\documentclass[epjST]{svjour}
\usepackage{amsmath}
\usepackage{graphics}
\usepackage{hyperref}
\usepackage[normalem]{ulem}
\usepackage{pdfpages}
\usepackage{siunitx}
\usepackage{textcomp}
\usepackage{soul}
\usepackage{color}
\definecolor{mauricio}{RGB}{21,168,48} 
\definecolor{nilton}{RGB}{153,153,0}
\definecolor{renan}{RGB}{128,0,128}
\definecolor{vinicius}{RGB}{150,75,30}
\begin{document}

\title{Building a model of the brain: from
detailed connectivity maps to network organization}

\author{
Renan Oliveira Shimoura\inst{1}\fnmsep\thanks{\email{renanshimoura@usp.br }} \and
Rodrigo F. O. Pena\inst{1,2} \and
Vinicius Lima\inst{1,3} \and
Nilton L. Kamiji\inst{1} \and
Mauricio Girardi-Schappo\inst{1,4} \and
Antonio C. Roque\inst{1}\fnmsep\thanks{\email{antonior@usp.br}}}
\institute{
Departamento de Física, FFCLRP, Universidade de São Paulo, Ribeirão Preto, SP, 14040-901, Brazil \and
Federated Department of Biological Sciences, New Jersey Institute of Technology and Rutgers University, Newark, New Jersey, NJ, USA \and
Institut de Neurosciences des Systèmes (INS, UMR 1106) Université de Aix-Marseille, Marseille, France \and
Department of Physics, University of Ottawa, Ottawa, ON, K1N 6N5, Canada}

%% abstract
\abstract{
The field of computational modeling of the brain is advancing so rapidly that now it is possible to model large scale networks representing different brain regions with a high level of biological detail in terms of numbers and synapses. For a theoretician approaching a neurobiological question, it is important to analyze the pros and cons of each of the models available. Here, we provide a tutorial review on recent models for different brain circuits, which are based on experimentally obtained connectivity maps. We discuss particularities that may be relevant to the modeler when choosing one of the reviewed models. The objective of this review is to give the reader a fair notion of the computational models covered, with emphasis on the corresponding connectivity maps, and how to use them.
} %end of abstract
\maketitle
\section{Introduction}
\label{intro}

\textcolor{blue}{With 86 billion neurons \cite{Azevedo2009} and hundreds of trillions of synapses \cite{BraSchuz1998}, the brain is one of the most complex systems in the known universe. Part of this complexity is due to the intricate pattern of connections among brain cells. Arguably, the computations performed by the brain depend heavily on the connections among its cells, but how? In other words, how the structure of the brain is related to its function? Many argue that since the brain is a complex system its functions are more than the sum of its parts \cite{Tononi1994,Tononi1998,Koch1999,BasGaz11}. So, knowledge of the individual behavior of the brain components is not enough to explain its emergent properties, e.g. cognition and consciousness. Therefore, the task of modeling the brain connectivity with a reasonable degree of accuracy constitutes an essential step for understanding these emergent phenomena.}

Building-block strategies where spatial and temporal scales are individually modeled are the \textit{modus operandi} of computational neuroscience. Ionic currents are modeled and studied in separate, neurons are studied in separate, populations are studied in separate, and the emerging behavior from the interaction of these ``blocks'' is then studied via mathematical and computational models. 
However, the step from single to multiple and interconnected cells is not trivial, neither from the point of view of behavior nor from the point of view of data extraction and coding.

But why is coding the structure of neuronal connections and not only the individual cells so important? We may look for hints on this question in different phenomena in nature. Starting with inanimate matter, we find that the crystalline structure of materials directly influences their thermal, electric, magnetic and optic properties~\cite{barkemaMC,gennesLC1993}. In addition, the dimensionality of the lattice, as expressed by the number of neighboring sites to each node in the network, as well as the reach of interactions, is known to alter significantly the behavior of physical observables in the vicinity of a phase transition~\cite{odorReview2004}. The flow of current and, consequently, the expected behavior of electric circuits, is highly dependent on the spatial configuration of resistors, sources, capacitors and inductors, and on how their branches intertwine between some input and output of electric signal. When conducting nanoparticles are arranged into a percolating structure, complex activity arises in the circuit via bursts of switching conductivity~\cite{mallinson2019}. Due to its intricate cortical-like activity, this condensed matter device could serve as prototype for neuromorphic hardware.

Although the brain is not an ideal and isolated electric circuit, the most accepted model for the neuronal membrane is based on an equivalent circuit made of a capacitor coupled in parallel to resistors and batteries~\cite{Johnston1995}. Hence, the structure of the brain, reflected on its immense electrical circuitry, works as a complex web that shapes neuronal activity, and ultimately determines brain function as a kind of process that lies in a continuous feedback loop with the embedding environment. Neuroscience is then tasked with relating a set of inputs to the brain (e.g., via sensory systems), to the corresponding generated outputs, also known as functions. Although the intended function is not always clear, we can have a look at the activity of specific brain regions to get clues on whether a particular structure of a computational model makes sense.

Examples of this structure-function interdependence come from experimental and theoretical studies alike~\cite{sporns2010,TomPen14,GirardiPLR2020}: spontaneous cortical activity is sometimes observed in the form of bursts of action potentials, known as neuronal avalanches~\cite{Hesse2014,Carvalho2021}. Nevertheless, when a group of researchers built a lattice-like network of neurons from the scratch \textit{in vitro}, they did not observe these complex patterns~\cite{Tibau2013}. However, adding modular structure to the lattice may do the job~\cite{yamamoto2018}. These modules could then be built on top of each other generating a
layered structure that mimics the cortex. These layered networks are known for generating propagating waves~\cite{girardiV1conf2015,muller2018cortical}, long-range correlations~\cite{Girardi2016,arnulfo2020long}, and neuronal avalanches~\cite{Girardi2018}. Rich-club-like modularity destroys bursting synchrony~\cite{lameu2012}, but allows for neuronal avalanches~\cite{Kaiser2007}, whereas hierarchical networks of this type optimize the diversity of spiking patterns~\cite{pena2020}. Some types of plasticity lead networks of neurons into a modular topology~\cite{lameu2019}. In fact, evolving techniques of manipulating neuronal activity may give birth to synthetic biological brain structures, also known as connectomes~\cite{rabinowitch2019would}.

While pointing to the importance of modeling the neuronal structure in a model, in this tutorial review we provide the reader a concise guide on how to access connectivity maps based on experimental recording, and from there how to model brain inspired networks. We discuss recent models using data-driven connectivity maps at neuronal level from different brain areas (e.g., cortex, hippocampus) and discuss advantages on choosing each of those models.

Our paper is organized as follows. The next section is divided in subsections where we present a brief discussion on how networks are usually constructed. We start in Section~\ref{sect:data} discussing how structural data is usually recorded from multiple experiments until it reaches the so-called connectivity map, passing through a discussion about basic neuron models in Section~\ref{sec:neuron_models} as well as synaptic models in Section~\ref{sec:synapse_models}. Then, in Section~\ref{sec:conn}, we guide the reader on how to code the connectivity maps into network models and describe what are the simulation tools that are commonly used. Finally, in Section~\ref{sec:summary} we summarize recent models based on data-driven structural connectivity maps, mainly the ones where connections are described at neuronal level as in microcircuit models.

\section{How network models are usually constructed}\label{sect:modeling}

With respect to connectivity, neural network models can be classified along two different axes. The first one refers to the nature of the graph used to implement the network, and the second to the granularity or scale of the connections.

Regarding the nature of the graph, it can be of two basic types:
\begin{itemize}
    \item \textbf{Artificial graph}. The connections among neurons are generated according to some predefined rules with the specific aim of creating a network with desired properties, e.g. random or small-world topologies~\cite{Roxin2004,Lin2005,sporns2010,BorPro17}. 
    \item \textbf{Data-driven graph}. The connections among neurons are based on experimental data obtained with different techniques with the aim of replicating as faithfully as possible the circuitry of a particular brain region~\cite{Bezaire2016,DuraBernal2017,Brunton2019}. 
\end{itemize}

Regarding the granularity of the connectivity, it is tied to the scale at which neural structures are described. There are three basic granularity levels:
\begin{itemize}
    \item Connections linking \textbf{morphologically detailed neurons}. Synaptic contacts happen at specific positions along cell bodies, e.g. distal/proximal dendrites or somata. Models that want to take into account the positions of synaptic contacts must be based on morphologically detailed neuron models. These models are composed of several interconnected compartments emulating the branched structure of the neuronal dendritic trees~\cite{Segev1998,Herz2006,sterratt2012}. Each compartment can have its own set of ionic channels and maximal ionic conductance densities. With this type of neuron model, connections among neurons can be set in a compartment-wise fashion, including compartment-specific values of the synaptic parameters.
    
    \item Connections linking \textbf{point neurons}. At a coarser grain level, neurons can be described as points without spatial structure. In such cases, to set the connections among neurons one needs only to specify which cells are connected to which (usually according to some probabilistic rules) together with the parameters (type and strength) of the cell-to-cell synapses~\cite{brunel2000,gerstner2014,Potjans2014}.
    
    \item Connections linking \textbf{neuronal populations}. At an even coarser spatial granularity level, individual cells are no longer recognized as such and groups of neurons are lumped together into single ``average'' neurons~\cite{DayAbb2001,Liley2015,Cowan2016}. In such neural population models the connections represent axonal links among neuronal groups or brain regions and the synaptic parameters correspond to effective properties of the existing synapses. 
\end{itemize}

The artificial graph approach has been used to study cortical activity states \cite{brunel2000,ostojic2014,pena2018,BorPro20}, specially transitions between up and down states \cite{destexhe2001,pena2018}; how basic information processing computations are performed by a population of neurons \cite{Vogels2005}; and to understand low-level operations performed by cell assemblies \cite{buzsaki2010neural,papadimitriou2020brain}. 

Noticeably, it is known that the primate cortex is organized in a structured manner \cite{sporns2005,Bullmore2011}. Indeed,  the connection among cortical regions resemble the structure of small-world networks \cite{watts1998collective}, with clusters sparsely connected among them but also with strong interconnections. There is also evidence that the cortex has a hierarchical structure \cite{mountcastle1997,kaiser2010,meunier2010}, meaning that the cluster has smaller clusters nested within them. This topology allows different regions to be relatively independent to process information and be specialized in distinct functions \cite{TomPen14,TomPen16,pena2020}. However, the existence of pathways connecting the clusters also allows for information to be integrated among different regions.

The data-driven approach is supported by the massive amount of data that is routinely gathered using different techniques on neuronal microcircuits of different brain regions~\cite{Shepherd2018}. They  demonstrate how microcircuits in the brain are highly specific in terms of their connectivity and the impact of this specificity on function. For example, there is a high specificity in the vertical pattern of connections among neurons in distinct cortical layers~\cite{ThoWes02,Binzegger2004}. The pattern of cortical microcircuitry endows the recurrent cortical network with specific computational properties~\cite{Li2013,lien2013}.  

In this tutorial review, we will focus on large-scale network models built using the data-driven approach. 
In the following sections we will explore the specificity of brain microcircuits while discussing modeling approaches based on maps that contain such information.

\subsection{Obtaining structural data}\label{sect:data}

In recent years, efforts to characterize the connectivity maps (connectomics) representing the cortical structure have been made \cite{sporns2011human,betzel2020network}. These connectivity maps can span several scales, from the intra- and interlayer connections in a cortical microcircuit to connections linking cortical regions \cite{PaxHua00,sporns2005,AliChu13,Ste13}, depending on the techniques used to obtain them. Since these maps track synaptic connections at the neuronal level, and white matter pathways connecting cortical areas at meso/macro-scale level, they are referred to as structural connectivity maps. Usually, the structural connections can be mapped using magnetic resonance imaging (MRI) based techniques such as diffusion tensor (or weighted) imaging (DTI/DWI), and tractography \cite{Basser1994}. The connectivity for local cortical microcircuits can be obtained by means of electrophysiological techniques \cite{Markram2015}, axonal tracing \cite{Kuypers1990,Saleeba2019}, and electron- \cite{Denk2004} and light- \cite{SheHar20} microscopy. More details about different ways to extract anatomical information and use them to build neuronal network models are given elsewhere \cite{VanAlbada2020}.

Similar approaches are employed for other brain areas. Anatomical explorations of the hippocampus date back to the works of Ramon y Cajal with Golgi staining techniques. The highly specific pattern of connections within the limbic system has been revealed through advanced MRI or neuroanatomical tract-tracing techniques  \cite{andersen2006,kajiwara2008,van2009,Witter2010,maller2019}.

With the increasing number of data collected, the structural or functional connectivity maps started to be used in modeling studies. Consequently, several projects aiming at the construction of realistic brain models, such as the Human Brain Project~\cite{Amunts2019}, the Blue Brain Project~\cite{Markram2006}, or the 
Allen Brain Explorer~\cite{Wang2020}. These are examples of projects being funded around the world \cite{kandel2013neuroscience,landhuis2017neuroscience}. Faithful models constructed upon the connectivity data available were built in order to understand how the brain connectivity structure impacts the dynamics of the system, and to which phenomena the information contained in these maps are crucial \cite{plesser2007efficient,Potjans2014,kunkel2014spiking,schmidt2015full}. A major advantage of such models is the construction of canonical models for similar brain areas: all over the neocortex there is a similar six-layered organization whereas for the archicortex and paleocortex there is a three- or four-layered structure ~\cite{She2017neo}. Once a specific microcircuit of such areas is built, it is then used as a building block to enlarge a given network model or to connect it to other areas. 

\subsection{The neuron model}\label{sec:neuron_models}

Although not the focus of this tutorial review, the choice of the neuron and synaptic models have an impact on the modeling results. So, they will be briefly reviewed in this and the next subsection. In terms of the neuron, the phenomenology of spike-train generation can be implemented without specific modeling of the underlying biophysical mechanisms but only by modeling the lipid bilayer of neurons by an equivalent passive RC circuit. In this simplified case the membrane voltage ($V$) is described by $\tau dV/dt = -V + RI = f(V)$,  where $\tau$ is the membrane time constant, $R$ is the membrane resistance, and $I$ the injected current. Since this model cannot generate an action potential by means of its own dynamics, an artificial mechanism of fire-and-reset is usually included where spikes are counted whenever $V$ crosses a certain threshold value $V_{\rm th}$ with a subsequent reset. The function of $V$ on the right-hand side is not restricted to be linear and can be nonlinear as well. These models are referred to as integrate-and-fire type models~\cite{gerstner2014}. 
    
The integrate-and-fire model can be extended in a variety of ways with the inclusion of elements that capture features of neuronal firing behavior. Two-dimensional integrate-and-fire models include equations to tackle feedback effects from ionic currents. Generally, such models can be expressed by the coupled ODEs: $dV/dt = f(V) - w$, $dw/dt = G(V,w)$. Due to their dynamical properties, two-dimensional integrate-and-fire models posses a richer repertoire of behaviors that can be fitted to experimental \textit{in vivo} and \textit{in vitro} data to reproduce characteristic firing patterns of neurons~\cite{Izhikevich2003,Brette2005}. Effects such as oscillations or subthreshold resonance can be described as well~\cite{pena2018b}.
Another class of neurons that can be used are the ones with discrete time, the
so-called map-based neurons~\cite{Girardi2013}. This class of neurons is particularly interesting
because they have a rich dynamical repertoire of spiking activity, and allow for relatively easy
analytical tractability and efficient simulations~\cite{Girardi2017}.

\begin{figure}[htp]
    \centering
    \includegraphics[scale=0.6]{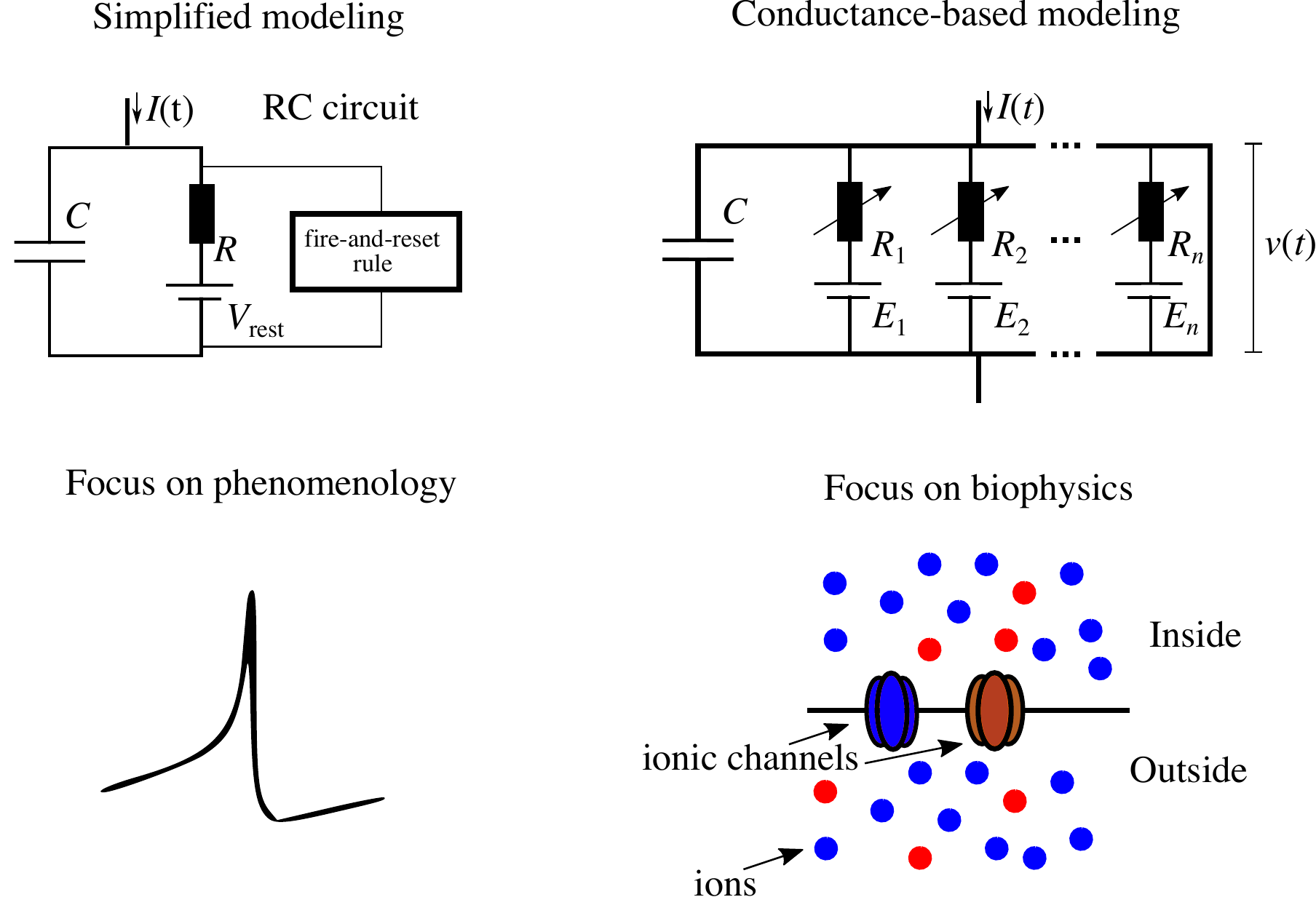}
    \caption{\textbf{Different types of neuron model}. Left: The simplified approach taken by integrate-and-fire models uses a fire-and-reset rule set ``by hand'' to model a spike. Right: The conductance-based approach models ionic currents using the Hodgkin-Huxley formalism.}
    \label{fig:neuron_models}
\end{figure}

Another class of neuron models is the one comprised by so-called conductance-based models, which explicitly describe the activation and inactivation dynamics of the gated ionic channels present in the neuronal membrane. This formalism originates from the seminal work of Hodgkin-Huxley in 1952~\cite{HodHux52}. The generic equations of a conductance-based neuron model are: $C dV/dt = - \sum_i I_i + I_{\rm inj}$, where $I_i = \bar{G}_i m_i^{r} h_i^{s} (V-E_i)$, and $dx_i/dt = (x_{i,\infty}(V) - x_i)/\tau_{x_{i}}(V)$ ($x = m$ or $h$). In these equations, $C$ is the membrane capacitance, $V$ is membrane voltage, $I_i$ is the ionic current of the $i$th ion, $\bar{G}_i$ is the maximal membrane conductance to ion $i$, $m_i$ and $h_i$ are, respectively, the activation and inactivation variables of the membrane conductance to the $i$th ion, $r$ and $s$ are small integers, $E_i$ is the reversal membrane potential of ion $i$, and $\tau_{x_{i}}(V)$ and $x_{i,\infty}(V)$ are, respectively, the voltage-dependent activation/inactivation time constant and steady-state value~\cite{sterratt2012}. In Fig.~\ref{fig:neuron_models} we show a schematic diagram the integrate-and-fire and conductance-based modeling approaches.
All these neuron models can be adapted to describe neurons with a determined morphology,
by defining discrete compartments (each of which obeys a certain membrane potential)
that are electrically coupled to each other.

\subsection{The synapse model} \label{sec:synapse_models}

Synapses can also be modeled at different levels of biological plausibility and the choice of synaptic model is crucial for the simulation of large-scale networks, since the number of synapses is much higher than the number of neurons. To model a synapse, one could simply define an event-driven model by incrementing the synaptic conductance or the synaptic current by a certain amount; or, on the other hand, consider more complex processes such as those of the conductance-based approach where the synaptic conductance depends on the membrane voltage of the postsynaptic neuron~\cite{Roth2010,gerstner2014}. Either way, the parameters of the model can be chosen to reproduce the behavior of excitatory synapses mediated by glutamate receptors AMPA and NMDA, and inhibitory synapses mediated by GABAergic receptors GABA$_{\rm A}$ (ionotropic) and GABA$_{\rm B}$ (metabotropic)~\cite{Roth2010}. Additionally, the synaptic strengths can be static or dynamic, and the latter case has been the focus of intense research aimed at modeling plastic synapses of both short and long term~\cite{TsoPaw98,castellani2001,tsodyks2005,clopath2010}.

The electric current $I_{ij}^{\rm syn}(t)$
generated by a single synapse from a neuron $j$ (presynaptic) to a neuron $i$ (postsynaptic) has the form~\cite{Roth2010},
\begin{equation}
\label{Eq:syncurr}
    I_{ij}^{\rm syn}(t) = G_{ij}(t) \left[V_i(t)-E_{ij}^{\rm syn}\right],
\end{equation}
where $G_{ij}(t)$ is a time-dependent conductance, $V_i(t)$ is the postsynaptic membrane potential, and $E_{ij}^{\rm syn}$ is the reversal potential of the synaptic current. The value of $E_{ij}^{\rm syn}$ determines whether the synapse $j\to i$ is inhibitory or excitatory (typical values of $E_{ij}^{\rm syn}$ are 0 mV for excitatory synapses and $-75$ mV for inhibitory ones).
For integrate-and-fire type models, a simplification that is often done is to fix $V_i$ (e.g. at resting voltage) and incorporate the battery term into $G_{ij}(t)$ so that Eq.~\eqref{Eq:syncurr} reads $I_{ij}^{\rm syn}(t) = J_{ij}(t)$. The term $J_{ij}(t)$ is the amplitude of the postsynaptic current (called synaptic strength or efficacy) and its sign determines whether the synapse is inhibitory (negative sign) or excitatory (positive sign).

\begin{figure}[htp!]
    \centering
    \includegraphics[scale=0.3]{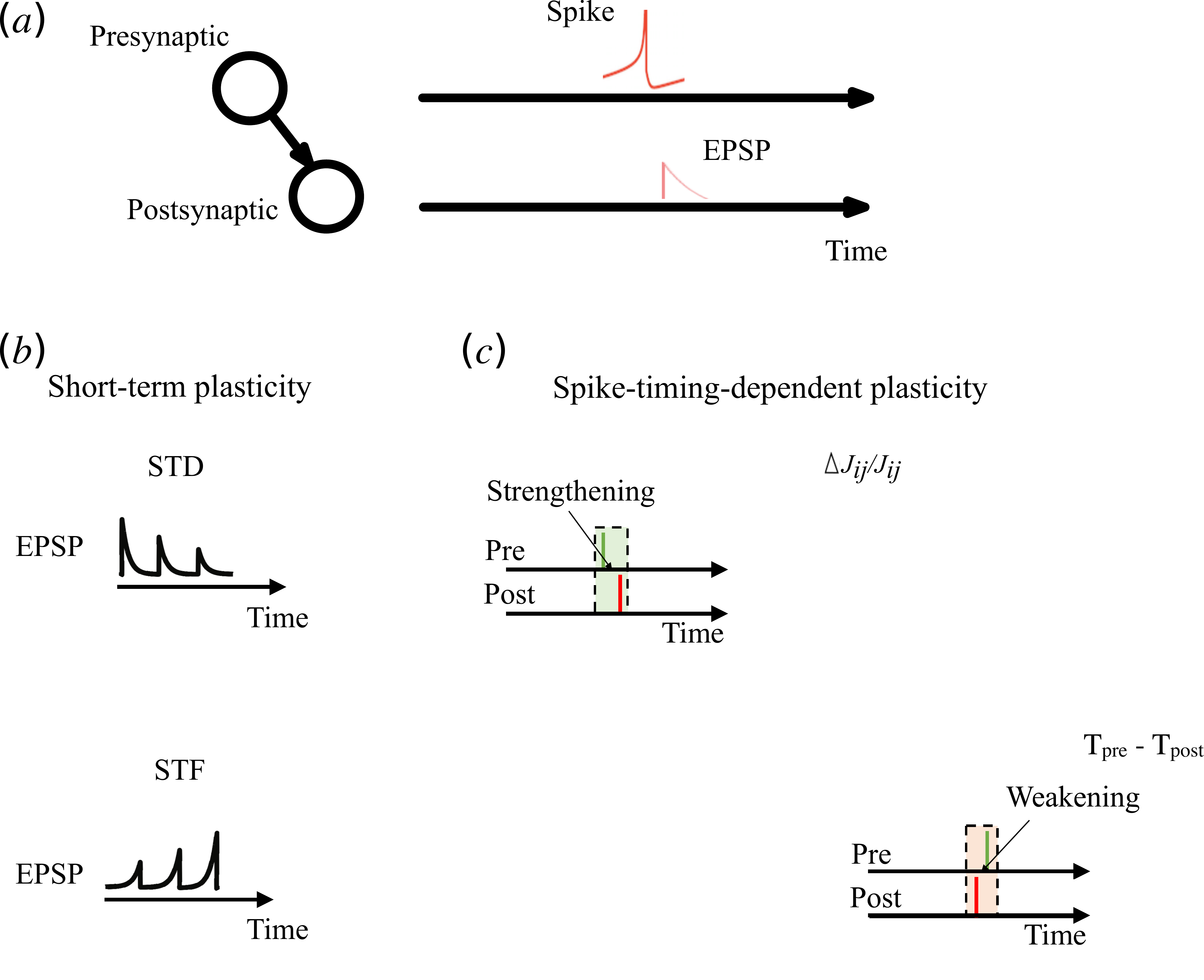}
    \caption{\textbf{Different synaptic models.} (a) Upon a presynaptic spike, an excitatory postsynaptic potential (EPSP) is created in the postsynaptic neuron after a synaptic time delay (a similar scheme can be implemented for an inhibitory synapse). The amplitude of the postsynaptic potential can change over time depending on some plasticity rule. (b) A short-term plasticity rule decreases or increases the EPSP amplitude over time depending on depression (STD) or facilitation (STF), respectively. (c) A spike-timing-dependent plasticity (STDP) rule changes the synaptic strength depending on the temporal difference between pre- and postsynaptic spikes.}
    \label{fig:sinapse}
\end{figure}

An example of short-term synaptic plasticity \cite{TsoPaw98} is the facilitation-depression dynamics given by Eqs~\eqref{Eq:stp1}--\eqref{Eq:stp}.
This model contains two variables that represent the fraction of presynaptic channels that are open $u_{ij}$, and the fraction of neurotransmitters that are available to be released $x_{ij}$. Upon a spike in the presynaptic neuron, the synaptic conductance $G_{ij}$ is increased by a factor $u_{ij}^{+} x_{ij} J_{ij} $, where $J_{ij}$ is the synaptic strength (a parameter) and the superscript $^+$ ($^-$) indicates the moment after (before) the spike. The interplay of the time constants $\tau_{\rm fac}$ and $\tau_{\rm dep}$ determines if a synapse will have a temporal depression or facilitation:
\begin{align}
\label{Eq:stp1}
\frac{d u_{ij}}{d t}= & -\frac{u_{ij}}{\tau_{\rm fac}}+U\left(1-u_{ij}^{-}\right) \delta\left(t-t^{f}\right), \\
\frac{d x_{ij}}{d t}= & \frac{1-x_{ij}}{\tau_{\rm dep}}-u_{ij}^{+} x \delta\left(t-t^{f}\right), \\
G_{ij}(t) = & u_{ij}^{+} x_{ij} J_{ij} \delta\left(t-t^{f}\right),
\label{Eq:stp}
\end{align}
where $t^f$ is the time at which the presynaptic neuron fires a spike,
and $U$ is the proportion of new open calcium channels upon a presynaptic event.
Examples of both short-term depression and facilitation can be seen in Fig.~\ref{fig:sinapse}.
The facilitation-depression model can be further simplified to a case without short-term 
plasticity by making $u$ and $x$ constants.

Alternatively, the synaptic strength can be determined by a spike timing dependent plasticity (STDP) rule, i.e., the increment or decrement of the synaptic efficacy is calculated using the relation between the times of pre- and postsynaptic spikes~\cite{MarLub97,MorDie08,sjo10}. Such rules are based on experimental evidence \cite{SjoPer01,shimoura2015}, and are applicable to both excitatory and inhibitory synapses \cite{KepVan02,KleFuk14}.
Let us assume that the synaptic efficacy of a $j\to i$ synapse can be described by a single variable $J_{ij}(t)$. The STDP rule can be implemented by defining two auxiliary variables, $x_j(t)$ and $y_i(t)$, which must be integrated over time. These variables are used to model the strengthening of the synapse when the presynaptic spike precedes the postsynaptic spike, and the weakening of the synapse when the presynaptic spike follows the postsynaptic spike, respectively. Then, a simplified STDP rule for excitatory synapses ($J_{ij} > 0$) is defined as~\cite{gerstner2014,sjo10}
\begin{align}
\label{Eq:stdp1}
\dfrac{dx_j}{dt} &= -\frac{x_j}{\tau_+} + \sum_f\delta(t-t^f_j)\:,\\
\label{Eq:stdp2}
\dfrac{dy_i}{dt} &= -\frac{y_i}{\tau_-} + \sum_f\delta(t-t^f_i)\:,\\
\label{Eq:stdp3}
\dfrac{dJ_{ij}}{dt} &= x_j(t)\,A_+\:{\rm\Theta}\!\left(J^{\rm max}-J_{ij}\right)\,\sum_f\delta(t-t^f_i)
-y_i(t)\,A_-\:{\rm\Theta}\!\left(J_{ij}\right)\,\sum_f\delta(t-t^f_j)\:,
\end{align}
where $t^f_j$ and $t^f_i$ are the spike times of the pre- and postsynaptic neurons, respectively, $\tau_{+,-}$ are the decay time constants of $x_j$ and $y_i$, respectively, ${\rm\Theta}(u)$ is the Heaviside function,
$A_{+,-}>0$ are the synaptic strength increment and decrement parameters, and $J^{\rm max}$ is the maximum value of synaptic efficacy.

Eqs.~\eqref{Eq:stdp1}
to~\eqref{Eq:stdp3} work like this:
every time a presynaptic spike occurs, two things happen:
first, $x_j$ is instantaneously increased by a unitary amount and then decreases exponentially with time constant $\tau_+$ and, second, while $J_{ij}>0$ it
is instantaneously decreased by $y_iA_-$.
On the other hand, every time a postsynaptic firing occurs, another two things happen: first, 
$y_i$ is instantaneously increased by an unitary amount and then decreases exponentially with time constant $\tau_-$ and, second, while $J_{ij}<J^{\rm max}$ it is instantaneously increased by $x_jA_+$.
This is captured by the scheme in Fig.~\ref{fig:sinapse}(c):
the variable $x_j$ is responsible for the exponential increase on the left-hand side
(green part) of the hyperbole,
since it only adds to $J_{ij}$ when $T_{\rm post}>T_{\rm pre}$ (a postsynaptic spike happens after a presynaptic spike); conversely,
$y_i$ is responsible for the exponential decrease on the right-hand side (red part) of the hyperbole, since it is only subtracted from
$J_{ij}$ when $T_{\rm post}<T_{\rm pre}$ (a postsynaptic spike happens before a presynaptic spike). Note that $T_{\rm post}=t^f_i$ and $T_{\rm pre}=t^f_j$ in the figure.

\subsection{How to implement connectivity maps in network models}
\label{sec:conn}

\subsubsection{Connectivity maps for networks of point neurons}

\begin{figure}[htp]
    \centering
    \includegraphics[scale=0.17]{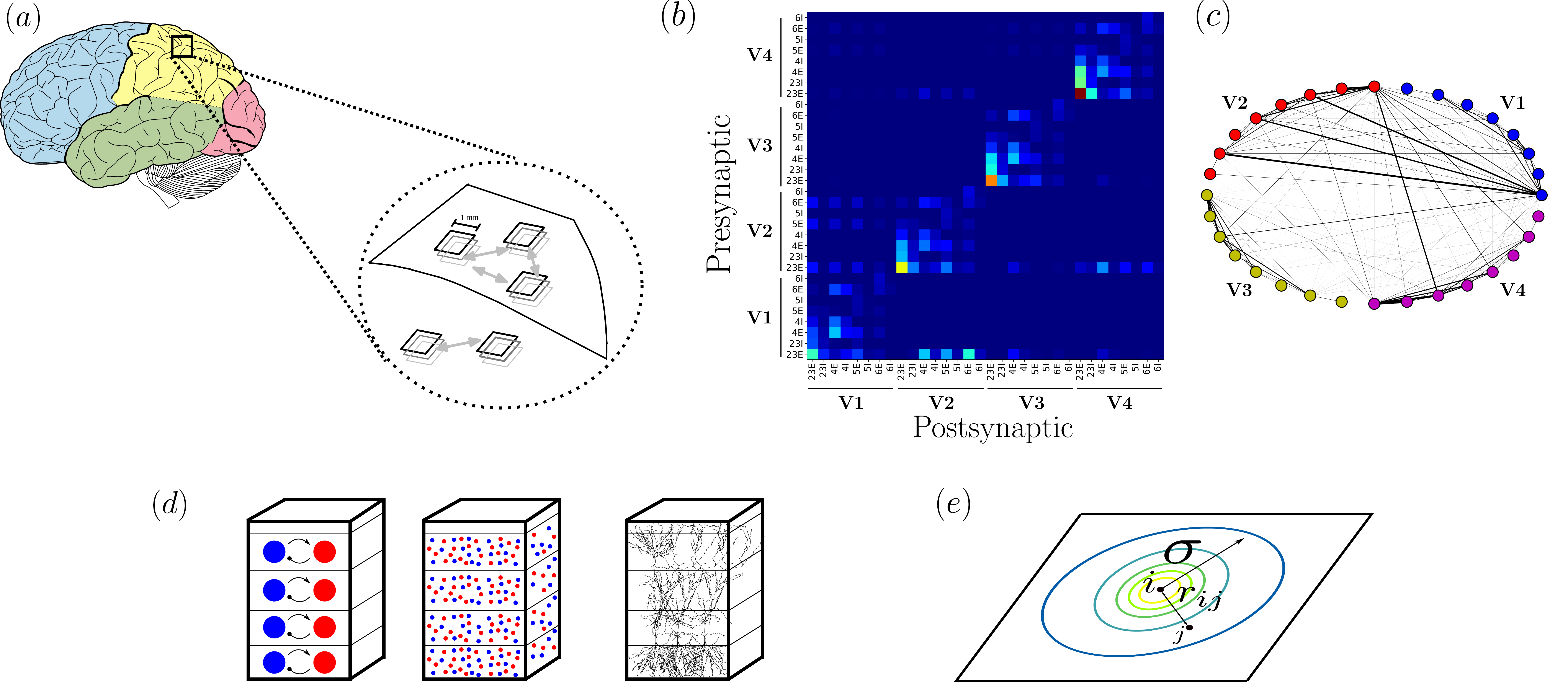}
    \caption{\textbf{From structural data to the neuronal network model.} Data for this example were taken from a multi-scale model of the macaque visual cortex \cite{Schmidt2018a,Schmidt2018b}. In (a) we show a illustrative depiction of the human cortex. The zoom indicates a cortical slice and the black squares represent cortical microcircuits. In (b) we show the structural connectivity matrix. Neurons are of excitatory or inhibitory types (E, I) and belong to four different cortical layers ($2/3$, $4$, $5$, and $6$) in four different cortical areas ($V1$, $V2$, $V3$, and $V4$). Presynaptic neurons are placed in the $x$-axis and postsynaptic neurons in the $y$-axis; Notice that the structural connectivity matrix is asymmetric. In (c) we show a network representation of the structural matrix in (b); neurons are placed along a ring and connections between pairs of them are indicated by lines. In (d) we show the different levels of spatial granularity at which a cortical microcircuit could be simulated: all excitatory/inhibitory neurons in a given layer can be described by a neural population model (left column); each individual cell can be represented by a point neuron model; or each individual cell can be described by a morphologically detailed neuron model (right column). As one goes from the left to the right column the number of equations and parameters of the full model increases dramatically, and, consequently, the computational cost involved in the implementation and simulation. This makes critical the choice of trade-off between the kind of phenomenon studied and the spatial granularity level of the model. In (e) we show how one could connect pairs of neurons using a distance based rule given by a probability density function as the 2D Gaussian function defined in Eq.~\ref{Eq:conn_gauss}. The colored circles indicate the contour lines of the distance based function and the presynaptic neuron is placed at the center of the inner contour line.}
    \label{fig:modelling_structure}
\end{figure}

Once with the data, the translation into a computational network model is not an easy task. Different levels of complexity require distinct ways to organize the extracted information into a connectivity map. Such a map may represent inter- or intra-areal connections, and in both cases the rules to implement the connections are similar. Our focus here is mainly on models at the microcircuit level, where a small piece of the brain as shown in Fig.~\ref{fig:modelling_structure}(a) is zoomed and its structure is represented by a connectivity matrix (Fig.~\ref{fig:modelling_structure}(b)) or ring of connections (Fig.~\ref{fig:modelling_structure}(c)). The same connectivity map can be used to generate networks with different levels of biological structural details as in Fig.~\ref{fig:modelling_structure}(d), where each node can represent a population of neurons or individual neurons. In the latter case, individual neurons can also be modeled by complex structures instead of just a point. Moreover, the information from the connectivity maps can also be expanded to add other features as spatiality (Fig.~\ref{fig:modelling_structure}(e)). In this section we will discuss in a guided manner how to better approach the above tasks.

Usually, connectivity maps are reported as matrices of connection probabilities where the rows/columns refer to the source/target elements (single neurons or neural populations) in the network. In a point neurons network with no spatial notion, the matrix gives all the information needed to build the network. The implementation is similar to the simplest case of a random network of the Erd\H{o}s-R{\'e}nyi type~\cite{Newman2010Book} where there is only one connection probability for all neuron pairs, thus one can draw connections by testing each neuron pair against this connection probability.

The way in which the testing mentioned above is implemented is of utmost importance. Choices such as pair combinations with or without replacement can create far different network graphs. Usually, the synaptic connectivity is established in a network model by attributing to each pair of neurons (or populations) a random number between zero and one drawn from an uniform distribution, and then testing it against the predefined connection probability for the pair. This choice avoids multiple synapses between the same neuron pair, which could be desirable or not. Moreover, data-driven connectivity maps at microcircuit level are usually directed graphs, i.e. pairs of neurons are not necessarily connected in a reciprocal way.

Alternatively, a distinct testing scheme would be to calculate the total number of synapses ($N_{\rm syn}$) among neurons based on the connection probability, and randomly draw lists containing $N_{\rm syn}$ pre- and postsynaptic indexes. In this method, multiple synapses are possible. An important detail is that these two schemes can deliver equivalent results depending on how $N_{\rm syn}$ is calculated.

An example of such differences can be found in a recent replication of the Potjans-Diesmann cortical microcircuit model \cite{Potjans2014} (reviewed in Section~\ref{sec:examples}) by us \cite{Shimoura2018}. Depending on the way the connection probability $C_{B,A}$ between a neuron $j$ in source population $A$ of size $N_A$ and a neuron $i$ in the target population $B$ of size $N_B$ is calculated, there are notable differences on the average spiking behavior of Layer 5 neurons. The exact value of $C_{B,A}$ is given by
\begin{equation}
\label{Eq:conn}
C_{B,A} = 1 - \left(1-\frac{1}{N_A  N_B}\right)^{N_{\rm syn}},
\end{equation}
\noindent where $N_{\rm syn}$ denotes the total number of synapses between populations $A$ and $B$. For $N_{syn}/(N_A N_B)$ small, the Taylor expansion of Eq.~\ref{Eq:conn} to first order results in the approximate expression for $C_{B,A}$ given by
\begin{equation}
\label{Eq:conn_approx}
C_{B,A} = \frac{N_{\rm syn}}{N_A \cdot N_B}.
\end{equation}

\begin{figure}[t!]
    \centering
    \includegraphics[scale=0.8]{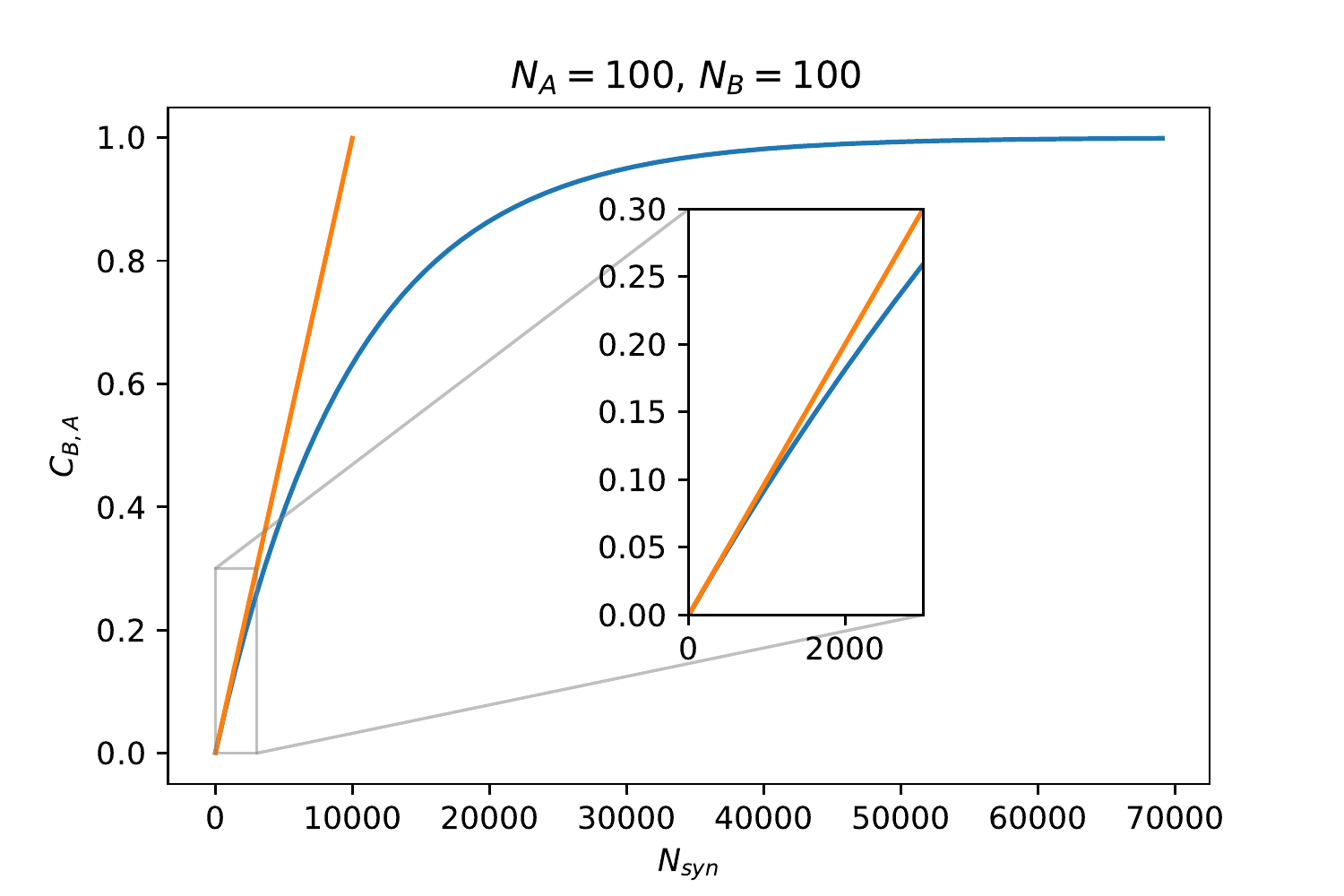}
    \caption{\textbf{Connection probability calculated by the exact expression and its first-order approximation.} Connection probability between two neuronal populations, $A$ and $B$, as a function of the number of synapses $N_{\rm syn}$ between them when calculated by the exact formula (Eq.~\ref{Eq:conn}) (blue) and its first-order approximation (Eq.~\ref{Eq:conn_approx}) (orange). Inset: zoom over small values of $N_{\rm syn}$ to highlight the beginning of the difference between the two curves.}
    \label{fig:conn_comparison}
\end{figure}

Fig.~\ref{fig:conn_comparison} shows a comparison of the curves of $C_{B,A}$ \emph{versus} $N_{\rm syn}$ calculated by Eqs.~\ref{Eq:conn} and~\ref{Eq:conn_approx} for fixed numbers of neurons in populations $A$ and $B$. For a small number of synapses $N_{\rm syn}$ the two equations give equivalent connection probabilities but as $N_{\rm syn}$ becomes larger the exact expression and its approximation diverge significantly. This reflects on the structure of the network and, consequently, on the activity (see \cite{Shimoura2018} for a more detailed discussion). 

Moreover, one has to determine whether connectivity maps indicate incoming or outgoing synapses. In a similar manner, instead of defining the total number of synapses between two populations, a distinction between in-degree and out-degree may be necessary. 

Another key point would be for maps that involve connectivity dependent on morphology. In these more complex structures, more elaborate methods involving multiple steps before deciding for the creation of connections may be necessary. Notice that the methods described above do not take into account the neuronal morphology or even the spatial notion while placing neurons on a grid. In order to incorporate these information, the connectivity matrix alone may not be sufficient.

From the point of view of spatial organization, there are limitations with respect to the maximum distance a neuron can send projections or to its target preferences. For the first case, a distance dependent connection probability may be required or, similarly, a fixed connection probability coupled to a distant dependent rule. Regardless of the way, neurons have to be placed on a spatial grid with a chosen dimension so distance dependent synapses can be created. As an example, consider the case of a two-dimensional (2D) grid where space is discretized in ($x$,$y$) positional variables and each neuron can assume a position in this ($x$,$y$)-grid. One can then define the absolute distance $r_{ij}$ between a presynaptic neuron $i$ and a postsynaptic neuron $j$,
\begin{align}
\label{Eq:distance}
r_{ij} = \sqrt{\Delta x_{ij}^2 + \Delta y_{ij}^2},
\end{align}
\noindent where $\Delta x_{ij} = \mid x_i - x_j\mid$ and $\Delta y_{ij} = \mid y_i - y_j\mid$. Notice that these distances may have limitations determined by boundary conditions. For example, in a 2D square grid of size $L \times L$ with periodic boundary conditions, $\Delta x_{ij}$ as defined above is only valid for $\mid x_i - x_j\mid \leq L/2$, otherwise $\Delta x_{ij} = L - \mid x_i - x_j\mid$. The same is valid for $\Delta y_{ij}$. Other boundary conditions could also be applied.

Once neuronal distances are handled, network connections can be set up by using the previously stated connectivity matrix as the zero-distance connection probability, and then test for pairs of neurons against a distance dependent connectivity rule. An example is given for a Gaussian probability density distribution,

\begin{align}
\label{Eq:conn_gauss}
c(r_{ij}) = C_{B,A} e^{-r_{ij}^2/2\sigma_A^2},
\end{align}

\noindent where $\sigma_A$ is the standard deviation. This equation is valid for $r_{ij} \leq R$, with $R$ the maximal distance a neuron projection can reach. A similar approach can be applied for a network with 3D spatial notion. Other distributions such as the exponential are often used.

\subsubsection{Connectivity maps for networks of neurons with morphology}

Other biological features can be implemented together with the distance dependence when creating the connections, an example is the direction tuning dependency for visual systems \cite{Billeh2020}. Nevertheless, in a more general manner, the next step towards adding structural complexity to the network is the implementation of neuronal morphology. For this, as an approximation, it is possible to use the same connection probability rules described above and define which pairs of neurons are connected using as reference their somata positions. Then, an additional procedure is required for creating the connections: the distribution of synapses along the neuronal dendritic tree. Different neurons may receive synaptic inputs with distinct distribution patterns in a cell-type and brain region dependent manner. 

Consideration of cell morphology is important when details related to specific locations of synaptic contacts are potentially relevant. An example is the organization of synaptic contacts involving inhibitory interneurons in the cortex. While parvalbumin-expressing (PV) interneurons target preferentially the somata of pyramidal cells, somatostatin-expressing (SOM) cells target their distal dendrites \cite{kawaguchi1997,tremblay2016}. This feature makes the response of a cortical pyramidal neuron to inhibitory input dependent on the type of cell that provides the inhibition. For example, the degree of attenuation of inhibitory postsynaptic potentials depends on where in the soma-dendritic domain of the pyramidal neuron they occur \cite{safari2017}. Similarly, in the hippocampus the modulatory effects on the spiking activity of pyramidal cells due to inhibitory inputs from oriens lacunosum moleculare (OLM) interneurons depend on the locations of these synaptic inputs \cite{Leao2012}. Brain networks are more than sets of nodes connected together and modellers must be aware that anatomical and morphological information could be key to a fuller understanding of brain functions.

\begin{figure}[htp]
    \centering
    \includegraphics[scale=0.6]{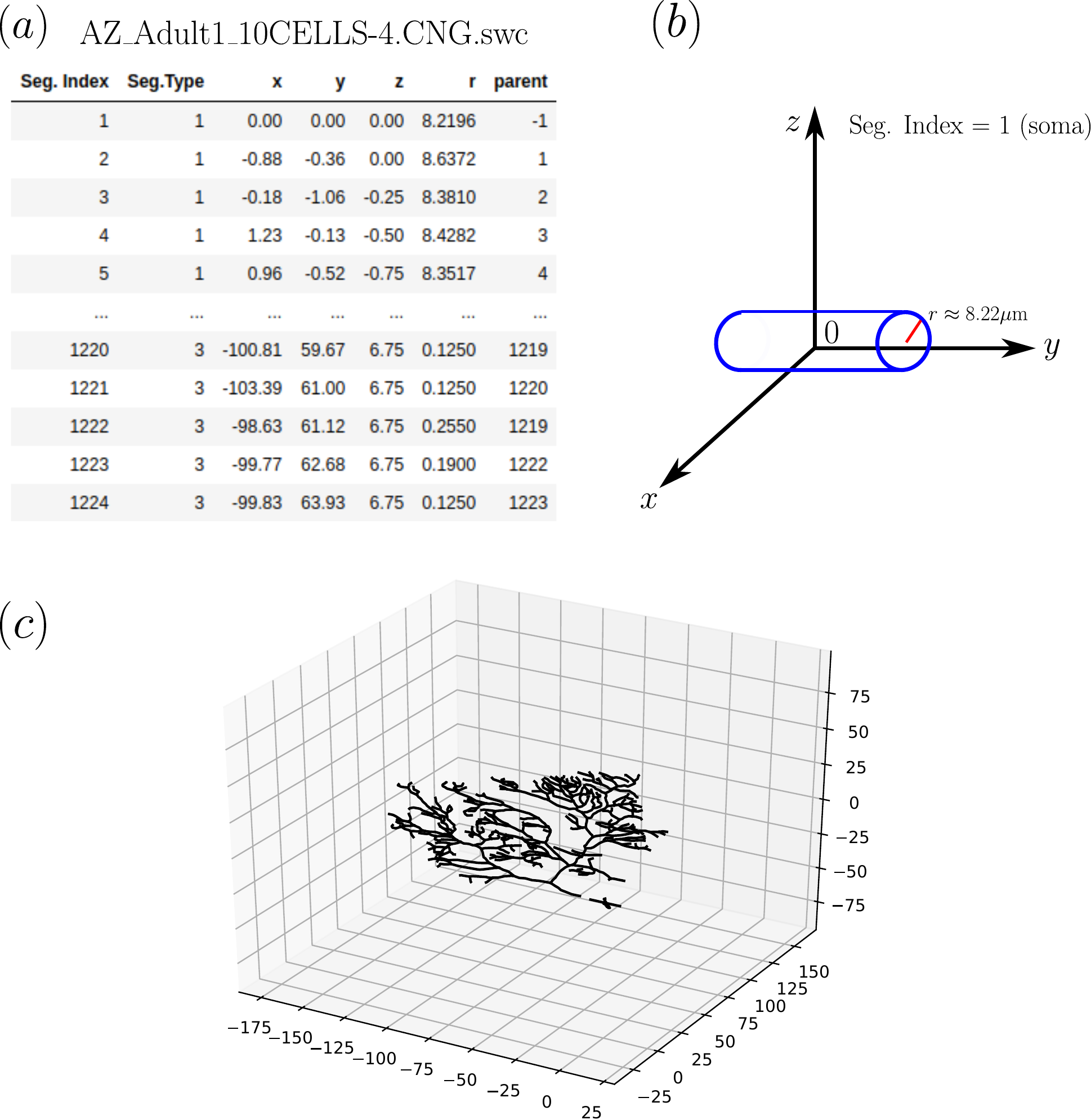}
    \caption{\textbf{Morphology using SWC files.} In (a) we show the content of the file AZ\_Adult1\_10CELLS-4.CNG.swc obtained from \href{neuromorpho.org/neuron_info.jsp?neuron_name=AZ\_Adult1\_10CELLS-4}{neuromorpho} corresponding to a mouse Purkinje cell. In (b) we depict the geometrical representation of the first segment (soma) from the same SWC file. In (c) we show a plot the cell morphology using the NEURON simulator.}
    \label{fig:morpho}
\end{figure}

Usually, the data-driven information needed to create the neuronal morphologies and the specific connection rules are given by the authors of the model or can be found on database repositories as \url{http://www.neuromorpho.org/}. Morphology files can be acquired by many different techniques and software, and it is common that files containing morphological information are in different formats such as Neurolucida \cite{glaser1990neuron}, SWC (baptized after the researchers who worked on a system to reconstruct the three-dimensional morphology of neurons \cite{stockley1993}), and MorphML \cite{crook2007morphml}. The morphological data available in the neuromorpho repository goes through a process of standardization before being published in order to make the data available uniform and easily readable across platforms. In Fig~\ref{fig:morpho}(a) we show the content of a SWC file corresponding to a Purkinje cell from the mouse. In each row the file displays the information of a given neuronal segment and the columns contain the following information from left to right: segment index, segment type, the coordinates x, y, z, the radius of the segment (in \textmu m), and the parent segment. While the interpretation of most of those properties is straightforward, the segment type and parent deserve further explanation. The segment type encodes which neuronal structure the segment represents as follows: soma = $1$, axon = $2$, (basal) dendrite = $3$, apical dendrite = $4$, and custom = $5+$. The first segment of the SWC file is always a soma. Notice that a given structure can be composed of many segments as the soma in Fig~\ref{fig:morpho}(a). The parent indicates the segment index $i$ at which the segment $j$ connects from, the first segment of the file must have parent value equal to $-1$, and for the subsequent segments the parent segment must always have a value smaller than that of the ``child'' segment.

Once the cell morphologies are implemented and the neurons are positioned in the spatial grid, one possibility for connecting them is to create connections between pairs of neurons according to the spatial intersections among presynaptic axons and postsynaptic dendrites. Additionally, the same distance-dependent connectivity rules discussed for point-neuron networks can be applied using the somata as reference points. It is important to note that processes (axon and dendrites) emanating from different neurons reach distinct distances. Some neurons make mostly local connections while other make both local and long-range contacts, and this can be taken into account in a distance-dependent way by setting distinct zones with predefined radii centered on reference points, e.g. somata, and allowing specific synaptic types to be created exclusively within such zones. After defining the existence of connections between pairs of neurons, the next step is to define the number of synapses and how they are distributed in their allocated zones. When there is not much information about synaptic positions, one strategy is to use a known distribution (e.g. uniform, Gaussian, etc.) and randomly place the synapses within the predefined zone of the postsynaptic dendritic tree until the maximum number of connections is attained. It is also known that specific neuron types tend to concentrate their received synapses on specific regions of their dendritic trees, and this information must be used to create a more biologically faithful connectivity pattern.

Since implementing these characteristics from the scratch in a code is a laborious task, these models are commonly implemented using neurosimulators. For morphologically detailed neuron models, the most used neurosimulator is the NEURON simulator \cite{Hines1997}. For instance, SWC files can be easily loaded into NEURON. Once imported, each segment is represented as a cylinder as depicted in Fig~\ref{fig:morpho}(b). The morphology can be visualized by means of the NEURON graphical user interface (GUI) or by calling the PlotShape function in a Python (or HOC, the original NEURON programming language) script. In Fig~\ref{fig:morpho}(c) we show a plot of the whole morphology of the Purkinje cell loaded from the SWC file. Besides NEURON, there are other tools to load cell morphologies. An example is NeuroConstruct \cite{gleeson2007neuroconstruct}, in which not only it is possible to load different types of morphology files but also to generate scripts compatible with different simulators such as NEURON, GENESIS \cite{bower2012book,Crone2019}, MOOSE \cite{Bhalla2008,Ray2008}, PSICS~\cite{cannon2010sto} and PyNN~\cite{davison2009pynn}. NEURON and other simulation environments will be better discussed in the next subsection.

Surely, there is not a unique method for building complex neuronal networks as different biological details may be modeled in distinct ways. In this subsection, we gave examples of general approaches to implement the connectivity map into a neuronal network code. It is important to realize that extracting synapses from a connectivity map derived from data is not a trivial task and may deliver different results depending on the approach. Many of the models that are discussed in the next section were built following the methods discussed in this section.

\subsubsection{Choosing a neurosimulator}\label{sec:neurosim}

Coding a model is a task that can be approached by low or high level programming languages. Many computational neuroscientists prefer to develop their own codes in low level languages such as C/C++, or Fortran. In higher level languages the syntax is more human readable and simpler to debug, as is the example of MATLAB (The MathWorks Inc., Natick, USA) which can be used as it is or coupled to packages for neuronal simulation \cite{heitmann2018brain}. While low level languages offer the advantage that the commands are closer to the processor instructions, dealing with complex data and complex syntax is not an easy task. Higher level languages ultimately make the models more accessible to the scientific community by using a more unified syntax, which facilitates information sharing and reproducibility~\cite{Nordlie2009,McDougal2016}. In this section, we present some of the most popular neurosimulators, which are high level packages developed with the sole purpose of neural modeling. We also discuss differences among them that should be taken into account when choosing the one to use.

In recent years, Python has become a standard programming language in many research areas due to its high productivity and interpretability. As a consequence, packages for many research fields such as astronomy \cite{robitaille2013astropy}, network analysis \cite{developers2010networkx}, machine learning \cite{pedregosa2011scikit,paszke2019pytorch}, and neuroscience \cite{muller2015python} are available for the scientific community. For computational neuroscience, in particular, many Python packages -- henceforth addressed to as neurosimulators -- can be used for the \textit{in silico} implementation of network models as the ones which will be reviewed in Section~\ref{sec:summary}.

We discuss three neurosimulators which are available in Python: Brian 2 \cite{stimberg2019brian}, NEST \cite{eppler2009pynest}, and NEURON \cite{hines2009neuron}. As discussed in Section~\ref{sec:neuron_models} the single neuron model can be classified as simplified or conductance-based, and can have morphology or not. When choosing a neurosimulator, one should first define in which of these categories the adopted model fits in.

The neurosimulator Brian 2 was developed to be used in Python~\cite{stimberg2019brian}. Although its main focus is on point neurons, Brian 2 also offers the possibility of defining cells with morphologies. A useful feature of Brian 2 is that it allows the user to define the ordinary differential equations (ODEs) of the model, so it is possible to describe both simplified or conductance-based models in the package. However, since the number of ODEs increases rapidly with the number of ionic channels modeled, Brian 2 is not very practical for detailed conductance-based models as the computational cost of implementing them increases rapidly. 

NEST~\cite{eppler2009pynest} constitutes another option for modeling large-scale networks of point neurons. Instead of allowing the user to define the ODEs of the model as in Brian 2, NEST has pre-implemented models available and the user only imports them. This is convenient for its practicality, but can be a problem if the model requires some mechanism that is not pre-implemented. For that, the user is either required to work with the NEST source code in C/C++ or use a software called NESTML which implements different models for NEST \cite{plotnikov2016nestml}. Both Brian 2 and NEST offer support to run the model in parallel, although only NEST offers message parsing interface (MPI) support. Because of that, NEST is more appropriate for large-scale models and is compatible with high-performance computing allowing a simulation to run across many compute nodes \cite{peyser2015nest,TikRub17}.

Finally, the NEURON simulator~\cite{hines2009neuron} is the alternative of choice for modeling neurons with morphology and networks made of them. In NEURON, a morphology is represented by a series of cylindrical compartments connected to each other, and even a single-compartment neuron with only a soma has a geometric representation with surface area and length. Even though one can adopt a simplified approach to model the compartment dynamics, the NEURON simulator works optimally with biophysical models where several ionic currents can be easily added to a neuron model. The simulator already has many ionic mechanisms implemented, such as the classic fast sodium and delayed rectifier potassium channels of the Hodgkin-Huxley model \cite{HodHux52} and many others, which can be found in databases such as modelDB \cite{hines2004modeldb}. If a specific ionic mechanism is needed, it can be programmed as a ``.mod'' file; however, this is a low level language and can be a nuisance for less experienced users. Despite the practicality of implementing morphological models in NEURON, building neuronal networks can be very challenging using this neurosimulator. In this regard, it is possible to use packages that work on top of NEURON such as NetPyne \cite{dura2019netpyne}, Brain Modeling Toolkit (BMTK) \cite{Dai2020b}, or BioNet \cite{GraBil18} to construct large-scale networks that can be efficiently run in parallel using NEURON. 

Many other packages for computational neuroscience are available, some of them, such as PyNN~\cite{davison2009pynn}, even try to integrate Brian 2, NEST, and NEURON. A complete characterization of the different neurosimulators available would be out of the scope of the present review; for this, we recommend the reader to see \cite{TikRub17,Blundell2018}. Several efforts have also been undertaken in recent years to promote code sharing and reproducibility in computational neuroscience \cite{Nordlie2009,Gutzen2018,McDougal2016,Mikowski2018,crook2020rep}. We emphasize the existence of repositories such as ModelDB \cite{McdMor17} and journals dedicated to the replication of computational work such as the ReScience Journal \cite{RouHin17}, which are important steps towards transparency in science.

\section{How modeling from connectivity maps can be done}\label{sec:summary}

As an example, one of the first challenging endeavors following a similar approach as described above was put forward by Izhikevich and Edelman~\cite{Izhikevich2008}. They proposed a hybrid model built from mixing diffusion tensor imaging (DTI) data with local circuitry information. The idea was basically to create a brain model from the scratch to check what type of collective behavior it exhibits, and learn on the fly what it takes to simulate a brain (as stated by one of the authors in his personal website~\cite{izhikevichsite}). There are many details to this model, and the authors give a comprehensive description of it in the supplementary material of their article. Here, we will describe only its main features. 

The entire cortex is reduced to fit into a sphere of 40 mm of diameter. This is done to keep the density of neurons appropriate. Neurons are randomly placed on the surface of the cortex, and they establish synaptic contacts according to a detailed map derived from two sources: first, the cortical surface map (that provides the coordinates for cortical neurons) was measured via anatomical MRI, whereas DTI provided the white matter tracts through which distant cortical regions are connected; this corresponds to the macroscopic structure of the model. Secondly, the microscopic structure (i.e., the cortical layers connectivity) came from a major study on cortical area 17 of the cat~\cite{Binzegger2004}, which yielded a long table of connection probabilities. It contained cortical layers 1 to 6 plus the thalamus and the reticular thalamic nucleus (RTN). The model also comprises many different neurons: pyramidal cells, spiny stellate neurons, basket and non-basket interneurons, thalamocortical relay cells, RTN cells and thalamic interneurons are some examples. These neurons have their morphologies partially described in terms of compartments representing the ramifications of their dendritic trees and axonal projections.

In regards to the dynamics, the model was originally built with 10$^6$ neurons and about 5$\times10^8$ synapses, although the precise number varied throughout the study. Dendritic compartments and somata are described by the two-dimensional quadratic integrate-and-fire model proposed by Izhikevich~\cite{Izhikevich2003}. The input current to each soma compartment is given by the sum of the dendritic current plus the synaptic current. The synaptic currents are composed of AMPA, NMDA, GABA$_{\rm A}$, GABA$_{\rm B}$ and gap junction contributions (the latter is only present inside layers and between specific sets of cells). The neurotransmitter-modulated synapses have time scale parameters and reversal potentials taken from the experimental literature, and their conductances obey first-order linear kinectics. The dendritic currents and the gap junctions are given by simple Ohmic currents. Short-term synaptic plasticity is modeled by a depressing or facilitating factor multiplying the synaptic conductances. Long-term plasticity is implemented via dendritic STDP rules using dopaminergic parameters, and the GABAergic synapses do not undergo these adaptations.

In terms of efficiency, it took the authors 1 minute to simulate approximately 1 second of network dynamics with sub-millisecond time step. In order to keep the network activity going on, the authors had to do a pre-run of the whole network subjected to spontaneous synaptic release (i.e., synaptic noise). Then, the synaptic plasticity present in the model was sufficient to generate a state with self-sustained activity. This state was used to detect signatures of chaos (sensitivity to initial conditions), brain rhythms, and traveling waves composed to up and down clusters of neurons with velocities consistent with experiments. Gamma rhythms were mostly present in the finer spatial scales, since averaging over a small piece of cortex hid these waves. Delta, beta and alpha rhythms also emerged. Interestingly, even though
the cortical microcircuitry was homogeneous, different regions of the cortex developed different rhythms -- this diversity might have come from the heterogeneity introduced by the long-range white matter tract connections.
The authors only suggested a way to calculate functional connectivity from their model, but did not do a thorough investigation to check what kinds of functional networks would appear.

The model by Izhikevich and Edelman~\cite{Izhikevich2008} is an examplar case of the modeling approaches and strategies described in the previous section. Decisions toward connectivity, neuron and synaptic models were made towards its construction and certainly influenced the final result. Note that the connectivity was primarily extracted from a single work \cite{Binzegger2004} following rules as described in Section~\ref{sec:conn}.

\subsection{Further examples of recent network models based on connectivity maps}\label{sec:examples}

Following the guidelines presented above, below we summarize different recent models which are entirely based on data-driven connectivity maps.  The list is far from being complete but illustrates to the reader how important areas of the brain can be approached in a similar manner. They can also be taken as a starting point by the newcomers to the field who want to pursue further studies in this field.

The models are listed in Tables~\ref{tab:references} and~\ref{tab:code_refs}. Table~\ref{tab:references} indicates the brain area modeled; the single neuron model used (simplified, which could be the one-dimensional leaky integrate-and-fire (LIF)~\cite{Tuckwell1988}, the two-dimensional Izhikevivh~\cite{Izhikevich2007} or adaptive exponential integrate-and-fire (AdEx)~\cite{Brette2005}, or the multivariable and multiparametric generalized integrate-and-fire (GLIF)~\cite{Teeter2018} models, or conductance-based (CB)~\cite{sterratt2012}); whether the neurons have morphology or are pointwise; the network size; whether the model implements synaptic plasticity or not; and whether the network has spatiality or not. Table~\ref{tab:code_refs} indicates whether the code of the model is publicly available or not and in which platform (see the web addresses indicated in the text), and the neurosimulator/programming language used. We want to point out that the code availability was based on the direct links presented explicitly in the references and, as some of the references are preprint papers, the code may be available after acceptance. Also, cases where the codes are available only by contacting the authors were not included.

\begin{table}[ht]
\centering
\caption{Summary of recent data-driven models with structural connectivity at microcircuit level.}
\label{tab:references}
\resizebox{\textwidth}{!}{%
\begin{tabular}{c|c|c|c|c|c|l}
\textbf{Brain area} & 
\textbf{\begin{tabular}[c]{@{}c@{}}Neuron\\ model\end{tabular}} & 
\textbf{Morphology} & \textbf{Network size} & \textbf{\begin{tabular}[c]{@{}c@{}}Synaptic\\ plasticity\end{tabular}} & \textbf{Spatiality} & \multicolumn{1}{c}{\textbf{Reference}} \\ \hline
Hippocampus CA1                       & CB         & Yes & 338,740               & No  & Yes & \cite{Bezaire2016} \\
Hippocampus CA1                       & CB         & Yes & 400,000               & Yes & Yes & \cite{Ecker2020} \\
Hippocampus                           & CB         & Yes & 112,200               & Yes & Yes & \cite{Hendrickson2012} \\
Olfactory bulb                        & CB         & Yes & 69,648                & Yes & Yes & \cite{Migliore2014} \\
Olfactory bulb                        & CB         & Yes & 97,652                & Yes & Yes & \cite{Migliore2015} \\
Primary motor cortex (M1)             & CB         & Yes & 10,000                & No  & Yes & \cite{DuraBernal2017} \\
Prefrontal cortex (PFC)               & AdEx       & No  & 1,000                 & Yes & No  & \cite{Hass2016} \\
Primary sensory cortices (V1, S1, A1) & LIF        & No  & 77,169                & No  & No  & \cite{Potjans2014} \\
Primary sensory cortices (V1, S1, A1) & LIF, CB    & Yes & 77,169                & No  & Yes & \cite{Hagen2016} \\
Primary sensory cortices (V1, S1, A1) & LIF        & No  & 1,249,136             & No  & Yes & \cite{Senk2018} \\
Visual cortical areas                 & LIF        & No  & 4,130,044             & No  & No  & \cite{Schmidt2018a,Schmidt2018b} \\
Visual areas (Retina, LGN, V1)        & Custom LIF & Simplified &$\sim$1,040,000 & No  & Yes & \cite{Girardi2016,Girardi2018,Girardi2016b}\\
V1                                    & CB, GLIF   & Yes & 230,924               & No  & Yes & \cite{Billeh2020} \\
V1                                    & CB         & Yes & 45,000                & No  & Yes & \cite{Arkhipov2018} \\
Primary somatosensory cortex (S1)     & CB         & Yes & $>$12,000             & No  & Yes & \cite{Reimann2013} \\
Primary somatosensory cortex (S1)     & CB         & Yes & 31,000                & No  & Yes & \cite{Markram2015,Ramaswamy2018,Nolte2019} \\
Striate cortex                        & AdEx       & Yes & 175,421               & No  & Yes & \cite{Tomsett2015} \\
Thalamocortical                       & Izhikevich & Yes & up to $100\times10^6$ & Yes & No  & \cite{Izhikevich2008} \\
\end{tabular}%
}
\end{table}

\begin{table}[ht]
\centering
\caption{Code availability and programming languages used in the reference works.}
\label{tab:code_refs}
\begin{tabular}{@{}c|c|l@{}}
\textbf{Code availability} & \textbf{Programming language} & \textbf{Reference} \\ \hline
\href{https://senselab.med.yale.edu/ModelDB/showModel.cshtml?model=187604\#tabs-1}{ModelDB}; \href{https://www.opensourcebrain.org/projects/nc\_ca1}{OSB} & NEURON & \cite{Bezaire2016} \\
- & Python, NEURON & \cite{Ecker2020} \\
- & Python, NEURON & \cite{Hendrickson2012} \\
\href{https://senselab.med.yale.edu/ModelDB/ShowModel?model=151681\#tabs-1}{ModelDB} & NEURON & \cite{Migliore2014} \\
\href{https://senselab.med.yale.edu/ModelDB/ShowModel?model=168591\#tabs-1}{ModelDB} & Python, NEURON & \cite{Migliore2015} \\
- & Python, NEURON, NetPyNE & \cite{DuraBernal2017} \\
\href{https://senselab.med.yale.edu/ModelDB/showmodel.cshtml?model=189160}{ModelDB} & C, Matlab & \cite{Hass2016} \\
\href{https://www.opensourcebrain.org/projects/potjansdiesmann2014}{OSB} & NEST & \cite{Potjans2014} \\
\href{https://github.com/INM-6/hybridLFPy}{GitHub} & Python, NEST, NEURON, LFPy, hybridLFPy & \cite{Hagen2016} \\
- & Python, NEST, NEURON, LFPy, hybridLFPy & \cite{Senk2018} \\
\href{https://github.com/INM-6/multi-area-model}{GitHub} & Python, NEST & \cite{Schmidt2018a,Schmidt2018b} \\
- & FORTRAN 95 & \cite{Girardi2016,Girardi2018,Girardi2016b}\\
\href{https://portal.brain-map.org/explore/models/mv1-all-layers}{Allen Brain Map} & Python, NEURON, NEST, BMTK & \cite{Billeh2020} \\
\href{https://github.com/AllenInstitute/sonata}{GitHub}; 
\href{https://github.com/AllenInstitute/arkhipov2018\_layer4}{GitHub} & Python, NEURON, BioNet & \cite{Arkhipov2018} \\
- & NEURON & \cite{Reimann2013} \\
\href{https://bbp.epfl.ch/nmc-portal/downloads}{Blue Brain Project} & Python, NEURON & \cite{Markram2015,Ramaswamy2018,Nolte2019} \\
\href{https://www.dynamic-connectome.org}{Dynamic Connectome} & Matlab & \cite{Tomsett2015} \\
- & C & \cite{Izhikevich2008} \\
\end{tabular}
\end{table}

\subsubsection{Models of the hippocampus}\label{sec:hippocampus_models}

The hippocampus is one of the most studied parts of the brain, and there are two basic reasons for this: first, it has a prominent laminar organization of neurons and their afferents, making it a popular model system for studies of cortical function; and second, it is involved with important functions as episodic memory formation, synaptic plasticity and spatial coding.
\newline

\noindent {\bf Interneuronal mechanisms of hippocampal theta oscillations in a full-scale model of the rodent CA1 circuit} \cite{Bezaire2016}: This is a full-scale (1:1) model of the rodent hippocampal CA1 region. The model has four layers with 338,740 neurons divided in 311,500 pyramidal cells and 27,240 interneurons. The pyramidal cells are described by the same multicompartmental conductance-based model with realistic morphology, and the interneurons are of eight different conductance-based types with simplified morphologies. Details of the neuronal dynamics including description of the ionic channels and their equations are given in the appendix section of the reference paper. The connectivity was extracted from experimental data and the synapses were modeled by a double exponential function without plasticity. To connect the neurons the distance between each pair of cells was taken into account together with the cell type dependent number of connections. Then the total incoming synapses to a postsynaptic neuron were divided into radial bins and distributed among the bins according to a Gaussian axonal bouton distribution of the presynaptic cell. The goal of the authors was to investigate theta oscillations and the importance of interneurons in the different phases. They found that parvalbumin-expressing interneurons and neurogliaform cells, as well as the interneuronal diversity itself is very important for the theta rhythm, although the gamma rhythm is also observed and discussed in the work. The code is available in ModelDB at \url{https://senselab.med.yale.edu/ModelDB/showModel.cshtml?model=187604}, and also in the open source brain (OSB) database at \url{https://www.opensourcebrain.org/projects/nc_ca1}.
\newline

\noindent {\bf Data-driven integration of hippocampal CA1 synaptic physiology in silico} \cite{Ecker2020}:
This is a detailed model of the rat hippocampal CA1 area following the same principles used in the construction of the model of the rat somatosensory cortex \cite{Markram2015} (see below). The model consists of about 400,000 cells ($\sim$90\% pyramidal and $\sim$10\% interneurons), represented by 11 morphologically reconstructed cell types described as multicompartmental conductance based models \cite{Migliore2018}. The model describes postsynaptic conductances with biexponential kinetics and has synaptic plasticity with STP dynamics described by the Tsodyks-Markram model \cite{TsodyksMarkram1997}. The model provides a complementary resource for the quantification of network structure in the rodent hippocampal CA1 region, and helps in the identification of gaps in existing knowledge.
\newline

\noindent {\bf Towards a large-scale biologically realistic model of the hippocampus} \cite{Hendrickson2012}: The model describes the pathway from layer II of the entorhinal cortex (EC) to the dentate gyrus (DG) (the  EC-to-DG pathway) of the rat hippocampus. This pathway comprises three pathways: EC to the CA3 region, DG to CA3, and CA3 to CA1. The model describes morphological structures such as dendritic trees based on experimentally measured parameters~\cite{Hillman1979}. The model is composed of $100,000$ granule cells, $11,200$ EC cells and $1,000$ basket cells. The connectivity map was extracted from experimental work~\cite{Yu2012}, which defines the presynaptic connections based on anatomical distances and then randomly distribute them until the spine pool is depleted. Synapses contain both STDP rules for long-term potentiation (LTP) and long-term depression (LTD) as well as STD and STP. Although the authors simulated only a reduced scale (to 1/10th) version of the full EC-to-DG pathway, their cluster simulations showed a good performance. Improvements of this model were done to study the topography dependency of spatio-temporal correlations in the Entorhinal-Dentate-CA3 circuit \cite{Yu2015}. A comparison between the 1/10th and the full scale versions of the model was implemented \cite{Hendrickson2015}. The code was developed in NEURON \cite{Carnevale2006,Hines1997} and the data analysis was made in Python.
\newline

\subsubsection{Models of the olfactory bulb}\label{sec:olfactory_bulb_models}
    
The olfactory system has a remarkable sensitivity and dynamic range, enabling the discrimination of thousands of volatile molecules (odorants) that occur in variable concentrations and move chaotically in the turbulent environment around us. This makes the olfactory system one of the favorite systems for the study of spatiotemporal representation and processing of stimuli in the brain. The olfactory bulb is the main structure in the olfactory processing pathway from nose to cortex, and because of this it has received considerable attention. There are many data available on the olfactory bulb and its afferent and efferent connections, allowing the development of computational models with high level of biological detail and connectivity information as the two examples given below.
\newline

\noindent {\bf Distributed organization of a brain microcircuit analyzed by three-dimen-sional modeling: the olfactory bulb} \cite{Migliore2014}: This model, which is based on a previous one-dimensional model of the olfactory bulb of the rat \cite{yu2013sparse}, is a three-dimensions model of the same structure constituted of cells with morphology and realistic input patterns. The main goal of the authors was to use the model to study the encoding of monomolecular odor stimuli on the mitral-granule cell circuitry. More specifically, they analyse this odor representation by inspecting the neuronal clusters formed between neurons due to synaptic plasticity after odor presentation. To constrain the peak conductances of the model to biological ranges, the authors first analysed the level of activity of $127$ glomeruli via optical imaging and used the normalized values in the model. The input is considered realistic because the peak conductance of the odor input is randomly drawn from a normal distribution given by optical imaging data. To model the morphology of mitral cells the authors used $8$ 3D experimentally obtained cell reconstructions. For each mitral cell the soma compartment was randomly drawn from one of these $8$ recorded morphologies, however the lateral dendrites were generated by an algorithm developed by the authors that generates the dendritic tree while preserving the following parameters from the experimental recordings: growth direction of the dendrites, path lengths, branch lengths, and the probability for each branch order. The authors generated a total of $635$ morphologies for mitral cells, which by their turn were connected to a variable number of glomerular cells ranging from 13,260 to 69,013 depending on the connectivity rule adopted. The background input to the granule cells, which was not explicitly modeled, consisted of Poisson generators. Synaptic plasticity (LTP and LTD rules) was implemented to model odor learning. Using the model, the authors showed that the glomerular pattern of activation is dependent on the odor input (in this study they used a set of $20$ natural odorant stimuli, e.g. coffee, kiwi and mint), suggesting that the glomerular activity patterns could be used by the system to encode the stimulus. The code is availabe in ModelDB at: \url{https://senselab.med.yale.edu/ModelDB/ShowModel?model=151681}.
\newline
     
\noindent {\bf Synaptic clusters function as odor operators in the olfactory bulb} \cite{Migliore2015}: The authors sought a better understanding on how odor stimuli is processed and interpreted by the mitral-granule circuit of the olfactory bulb by analyzing the configuration of the synaptic weights, referred to as synaptic cluster, after stimulus presentations. The cluster formed by a single glomerular unit was defined as a column. The model, as described above~\cite{Migliore2014}, consists of a 3D model, which enables a realistic representation of the overlapping neuronal dendritic trees. It consists of 635 mitral cells organized in 127 glomeruli (5 per glomerulus), and 97,017 granule cells. The network has synaptic plasticity (LTP and LTD), and the weights are modified independently in each dendrodendritic synapse between mitral and granule cells, depending on the local spiking activity, after the presentation of an odorant input. The model was validated by comparing the formation of synaptic clusters after stimulus presentation with experimental data representing a cluster configuration, obtained from pseudorabies virus staining. Among the main simulation results reported by the authors, we can highlight: (i) the model reproduces the spread of backpropagated action potentials originated in the dendrites of mitral cells within a glomerulus as observed experimentally; (ii) an analysis of the final configuration of synaptic weights after exposition to two odorant stimuli in different sequences showed that odor exposure is noncomutative, suggesting that the olfactory bulb represents odors learned in the past depending on the order they were learned. The authors also developed a mathematical framework to describe how the circuit represents the outputs as a function based on a series of matrix operations. The code is availabe in ModelDB at: \url{https://senselab.med.yale.edu/ModelDB/ShowModel?model=168591}.
\newline

\subsubsection{Models of frontal lobe regions}\label{sec:frontal_lobe_model}

The frontal lobes comprise about one-third of the cortical tissue. This large area is involved in a wide range of motor and executive functions, including simple and complex motor skills, attention, judgement, reasoning, problem solving and emotional regulation. The frontal lobes are also extensively connected with other brain regions. The complexity of the structures and functions of the frontal lobes has hindered our understanding of how these structures and functions are related. Modeling is likely to play a key role in future advances on this theme. We present below two examples of models for frontal lobe regions: the primary motor cortex and the prefrontal cortex.
\newline

\noindent {\bf Multiscale dynamics and information flow in a data-driven model of the primary motor cortex microcircuit} \cite{DuraBernal2017}: This is a biophysically detailed model of the mouse primary motor cortex (M1) with over 10,000 neurons and 35 million synapses. The modeled M1 microcircuit is represented by a cylindrical volume of $300 \mu m$ of diameter divided into six layers (L1, L2/3, L4, L5A, L5B and L6). Seven different types of pyramidal cells and two inhibitory interneurons were implemented, all using a conductance-based formalism. Among the different types of neurons, the authors focused and gave more biological details to two neurons from layer 5: the intratelencephalic (IT) and the pyramidal-tract (PT) neurons. The IT and PT neurons are modeled with full dendritic morphologies and have $700+$ compartments. For the other cells, simpler models were used containing around $3-6$ compartments. The synapses are conductance-based and not plastic representing AMPA and NMDA for excitatory connections, and $GABA_{\rm A}$ and $GABA_{\rm B}$ for inhibitory connections. The model exhibits spontaneous neural activity patterns and oscillations consistent with M1 data, which are dependent on the cell class and the layer and sublaminar location. The simulated local field potential (LFP) displays delta and beta/gamma oscillations with gamma amplitude modulated by delta phase. The flow of information measured by spectral Granger causality indicates routes at frequencies in the high beta/low gamma band. Furthermore, the brief activation of specific long-range inputs or different neuromodulatory conditions changes the output dynamics. The results of the model simulations suggest an association between the cell-type-specific circuits of M1 and dynamic aspects of activity, such as oscillations, neuromodulation and information flow. The model was developed using NEURON~\cite{Carnevale2006,Hines1997} and NetPyNE~\cite{dura2019netpyne}.
\newline

\noindent {\bf A detailed data-driven network model of prefrontal cortex reproduces key features of in vivo activity} \cite{Hass2016}: The model is a detailed neuronal network constructed after \emph{in vitro} electrophysiological and anatomical data from the prefrontal cortex (PFC) of the rat and mouse. The 1,000 point neurons in the model are described by a simplified version of the AdEx model (simpAdEx \cite{Hertag2012}). Population heterogeneity is taken into account by the four different neuron types used in the network: pyramidal cell, fast-spiking interneuron, bitufted interneuron, and Martinotti cell. These neurons are distributed into two cortical layers, representing the superficial layers 2/3 and the deep layer 5. The connections are randomly created following different connection probabilities for each pair of cell types. Missing experimental information about rodents was replaced by data on monkeys, ferrets and cats. Synapses (AMPA, NMDA and GABA$_{\rm A}$) are conductance-based modeled by double-exponential functions. The model also includes short-term plasticity dynamics. Different characteristics such as spike train statistics, membrane potential fluctuations, local field potentials, and transmission of transient stimulus information across layers were validated by the authors against experimental recordings, and the validation appears robust even with moderate changes in the parameters. The code, written in C and MATLAB (The MathWorks Inc., Natick, USA), is available at \url{https://senselab.med.yale.edu/ModelDB/showmodel.cshtml?model=189160}.

\subsubsection{Models of sensory cortices}\label{sec:sensory_cortex_model}

The sensory cortices are responsible for receiving and interpreting sensory information from different modalities. The similarities experimentally observed among the local pattern of connections across different sensory cortices of different animal species led to the concept of ``canonical neocortical microcircuit'' \cite{Douglas1989}. This idea is implemented in the different computational models presented in this subsection, which focus on the somatosensory and visual systems. The examples show implementations with different levels of biological detail, from networks of point neurons to complex networks of neurons with morphology and synaptic plasticity. They serve as an illustration of how, once with a connectivity map, the level of biological detail put in the model may vary depending on the scientific question. Some of the examples below use the same connectivity map to generate networks with different mechanisms implemented at neuronal and/or synaptic level.
\newline

\noindent {\bf The cell-type specific cortical microcircuit: Relating structure and activity in a full-Scale spiking network model} \cite{Potjans2014}: This is a multilayered model of the microcircuit of the early sensory cortex based on data from primary sensory cortices of the cat, rat, and mouse. The model combines anatomical and physiological data in a detailed implementation of the canonical neocortical microcircuit~\cite{Douglas1989}. The missing experimental data was replaced by estimates and approximations involving target specificity. The model reproduces in full-scale the neuronal network under a surface area of 1 mm$^2$ of early sensory cortex, and contains 77,169 excitatory and inhibitory neurons divided into four layers, comprising eight cell populations: L2/3e, L2/3i, L4e, L4i, L5e, L5i, L6e, and L6i. All neurons are point neurons described by the same leaky integrate-and-fire model. The network is built from a given connectivity matrix with probabilities of connection between each pair of populations. With short delays after presynaptic spikes, synaptic currents receive step increases and then decay exponentially. The step magnitudes and decay time constants are fixed and depend on the synapse type (excitatory or inhibitory). Synaptic delays are randomly drawn from an uniform distribution. Inputs coming from other brain regions are modeled by Poisson spike trains. The model generates spontaneous asynchronous irregular activity and cell-type specific firing rates in good agreement with \emph{in vivo} recordings. The code is available in the OSB web site at \url{http://opensourcebrain.org/projects/potjansdiesmann2014} (versions in NEST, PyNEST, and PyNN). A replication of the model in Brian 2~\cite{Shimoura2018} is available at  \url{https://github.com/ReScience-Archives/ShimouraR-KamijiNL-PenaRFO-CordeiroVL-CeballosCC-RomaroC-RoqueAC-2017/}.
\newline

\noindent {\bf Hybrid scheme for modeling local field potentials from point-neuron networks} \cite{Hagen2016}: This work proposes a hybrid modeling scheme combining point-network models with biophysical principles underlying LFP generation by neurons. To demonstrate the hybrid scheme the authors use two models: the first is a modified version of the point-neuron network of Potjans and Diesmann described above~\cite{Potjans2014}. This network is simulated and its spiking activity is recorded. The second model is a network of multicompartmental neurons which receive cell-type and layer-specific synaptic inputs coming from the spikes generated by the first model. The multicompartmental neurons belong to 16 different cell classes reconstructed from several published sources and mainly from the cat visual cortex. Similarly to the model by Izhikevich and Edelman \cite{Izhikevich2008}, the fractions of neuron types per layer were extracted and adapted to create the multicompartment version, which was used to estimate the LFP.
\newline

\noindent {\bf Reconciliation of weak pairwise spike-train correlations and highly coherent local field potentials across space} \cite{Senk2018}: This model is an extension of Potjans and Diesmann model \cite{Potjans2014} to include a spatial notion and an upscale in the surface area delimiting the cortical column from 1 mm$^2$ to $4 \times 4$ mm$^2$. The network size was chosen to match the size of the Utah multi-electrode array and to reproduce LFP measurements related to spatial propagation and distance-dependent correlation of evoked neural activity. The number of neurons, totaling 1,249,136, was adjusted to keep the same neuron density by layer as the Potjans and Diesmann model, and the connectivity matrix was modified resulting in a matrix of zero-distance connection probabilities. The point neurons are modeled by the LIF equations and are randomly placed in a 2D square grid representing the same size of the surface area. The synaptic connections are randomly created by a distance-based connection probability rule following a Gaussian distribution. The network activity preserves the main characteristics of the Potjans and Diesmann model, including the asynchronous irregular spiking across populations. Moreover, in agreement with what is experimentally observed in the sensory cortex, the model displays weak pairwise spike-train correlations while also showing strong and distance-dependent LFP correlations. The network is implemented in NEST \cite{eppler2009pynest} and the LFP is measured using NEURON, LFPy and hybridLFPy. The data analysis in the article was done using Python.
\newline
    
\noindent {\bf Multi-scale account of the network structure of macaque visual cortex} \cite{Schmidt2018a}: This model is an expansion of the Potjans and Diesmann model \cite{Potjans2014} from a single microcircuit to a multi-area network representing the 32 areas related to vision in the cortex of the macaque. Each one of the 32 areas is constituted by a 1 mm$^2$ microcircuit containing four layers (except the parahippocampal area which has only three layers) with populations of excitatory and inhibitory neurons as in the Potjans and Diesmann model. The number of neurons by area vary from 197,936 (area V1) to 73,251 neurons (area TH). The neuron and synaptic models are the same as used in the Potjans and Diesmann model. There are adjustments in the neuron density and layer thickness based on the region where the area is located. The long-range connections (among areas) reproduce layer-specific inter-area projections taken from experimental data. The authors describe in the article all the steps they used to obtain the structural map for local and long-range connections, including approximations and references. Although the connectivity matrix is not explicitly shown, in the supplementary data there are tables with all values required to built the model. The code is available in GitHub at \url{https://github.com/INM-6/multi-area-model}, and from that it is possible to extract all the necessary information to reimplement the model. 
\newline
    
\noindent {\bf A multi-scale layer-resolved spiking network model of resting-state dynamics in macaque visual cortical areas} \cite{Schmidt2018b}: In this work the authors studied the dynamics of the multi-area model described above \cite{Schmidt2018a}. The network reproduces spiking statistics from electrophysiological recordings and cortico-cortical functional connectivity patterns observed in functional magnetic resonance imaging (fMRI) studies under resting-state conditions.
\newline

\noindent {\bf Griffiths phase and long-range correlations in a biologically motivated visual cortex model} \cite{Girardi2016}: This is a detailed network model representing the form recognition pathway of the visual system of mammals. It comprises five layers: the retina photoreceptors, the thalamic lateral geniculate nucleus (LGN), and the cortical layers 4C$\beta$, 2/3, and 6. Neurons are modeled by
a discrete-time compartmental integrate-and-fire dynamics with dissipative dendritic compartments and conservative axon and soma compartments. The number of dendritic and axonal compartments is adjusted to keep the velocity of propagation of the action potential comparable to experiments. The neurons have a simplified morphology, where dendritic, soma, and axonal compartments are organized in a linear structure. The structural matrix for interlayer excitatory connectivity is given by the authors, and was drawn from experiments with the macaque monkey. Although lateral inhibition is not taken into account, the network excitation is balanced by dissipation in the dendrites. The retina layer contains $10^6$ photoreceptors, and the other layers contain around $100^2$ neurons each, summing up to 40,000 neurons
in the network. A total of up to $32.5\times10^6$ synapses are formed by linking axonal to dendritic compartments according to predefined distributions of synaptic boutons over the bodies of the neurons.
The network has recurrent connections between layers 2/3 and 4C$\beta$, and interlayer connections are made according to a Gaussian distribution around the presynaptic neuron position, creating a columnar structure. The authors use the size of postsynaptic potential (PSP), obtained as an average over macaque brain experiments, as a control parameter, and the network starts to display activity in a complex feedback pattern as the PSP is increased past 1~mV~\cite{Girardi2016b}. The phase transition happens via an unusual type of nonequilibrium percolation over a Griffiths region~\cite{Girardi2018}, when compared to the usual directed-percolation frequently found in neuronal networks~\cite{Carvalho2021,Girardi2019,Girardi2020bal,Girardi2021theory}. The network also presents power-law avalanches and long-range
correlations (with approximately $1/f$ power spectrum of avalanche sizes)~\cite{Girardi2016,Girardi2018}.
\newline
    
\noindent {\bf Systematic integration of structural and functional data into multi-scale models of mouse primary visual cortex} \cite{Billeh2020}: This is a data-driven model of the mouse primary visual cortex (V1), containing $230,924$ neurons that simulates physiological studies with arbitrary visual stimuli. The model contains 17 cell classes divided into five cortical layers (1, 2/3, 4, 5 and 6) and represents a cortical column with \SI{845}{\micro\metre} of radius. The model contains a connectivity matrix giving the connection probabilities between all neurons types at each layer. Synapses are modeled by bi-exponential functions and are not plastic. Two variants of the model were developed, a version with biophysically detailed compartmental neuronal models \cite{Gouwens2018}, and a version with generalized leaky integrate and fire (GLIF) point-neuron models \cite{Teeter2018}. Both versions are constrained by experimental measurements and reproduce many observations from electrical recordings \emph{in vivo} \cite{Siegle2019}. Also, in both models the firing rate distributions in response to the used inputs are similar. The authors say that the tuning adjustments made in the networks to reproduce experimental data were useful to identify rules governing cell-class-specific and synaptic strengths. The simulations were developed with the use of the Brain Modeling ToolKit (BMTK) (https://alleninstitute.github.io/bmtk, \cite{Gratiy2018,Dai2020b}), which facilitates simulations with both NEURON \cite{Hines1997} and NEST \cite{Gewaltig2007} and supports Python 2.7 and 3.6. Model and simulation outputs are saved in the SONATA format (\url{https://github.com/AllenInstitute/sonata} \cite{Dai2020}). All models, code, and meta-data are publicly available via the Allen Brain Map web portal at \url{https://portal.brain-map.org/explore/models/mv1-all-layers}.
\newline
    
\noindent {\bf Visual physiology of the layer 4 cortical circuit in silico} \cite{Arkhipov2018}: This is a data-driven biophysically and anatomically detailed circuit model of layer 4 from the primary visual cortex (V1) of the mouse. The detailed network has $10,000$ neurons divided into five cell types, three excitatory and two inhibitory, which are modeled using individual neuron parameters from the Allen Cell Types Database (\url{http://celltypes.brain-map.org/}). The connectivity map is based on data from V1 of the cat. To prevent boundary artefacts, the detailed model was surrounded by a simplified network of $35,000$ leaky integrate-and-fire neurons divided into excitatory and inhibitory groups. The cells were spatially distributed and the connections were randomly created according to a probability rule that implements linear decay with intersomatic distance and dependence on the difference between the assigned preferred orientation angle. Once a connection was established, the number of synapses is randomly selected with uniform distribution in the range between 3 and 7. Additionally, a functional synaptic connectivity rule, commonly observed in the excitatory neurons in the mouse V1 L2/3 and known as ``like-to-like connectivity'', is applied. The positions of the synapses along a postsynaptic dendritic tree are randomly assigned with distance constraints based on experimental data. The adopted synaptic function models are the bi-exponential for the biophysical neurons and the instantaneous rise followed by exponential decay for the LIF cells. The inputs coming from the lateral geniculate nucleus (LGN) of the thalamus are simulated as a set of filters representing the image processing done at pre-cortical stages. The same or similar sets of visual stimuli presented to the model was also presented for comparison to mice in \emph{in vivo} experiments. The model reproduces a large set of experimental results, including effects of optogenetic perturbations. Simulations showed that tuning properties of neurons are affected by the functional connectivity rules. A simplified version of the network was simulated and the results were qualitatively similar but lacked quantitative agreement. The simulations were done using Python 2.7 together with NEURON 7.4 \cite{Carnevale2006} through the BioNet package \cite{Gratiy2018}, which is an interface for modelling large-scale networks. The codes are available at: \url{https://github.com/AllenInstitute/arkhipov2018_layer4}.
\newline
    
\noindent {\bf A biophysically detailed model of neocortical local field potentials predicts the critical role of active membrane currents} \cite{Reimann2013}: This is a model of cortical layers 4 and 5 populated by over $12,000$ multicompartmental pyramidal and basket cells. The data used comes from the rat primary somatosensory barrel cortex (S1). There are five million dendritic and somatic compartments with voltage- and ion-dependent currents, realistic connectivity, and probabilistic AMPA, NMDA, and GABA synapses. The neurons somata were randomly placed, according to the layer depth, in a 3D hexagonal volume with radius of \SI{320}{\micro\metre} and the synaptic connections were created for $5 \%$ of apposition closer than \SI{3}{\micro\metre}. Simulation results show that the LFP is dominated by active currents instead of synaptic currents. This is a convenient model for LFP reproduction and to study how it is affect by internal and external layer interactions in cortical circuits. The model is part of a collaborative effort between several institutions including the Blue Brain Project and the Allen Institute for Neuroscience. The code, unfortunately not available, is written in NEURON \cite{Hines1997}.
\newline
    
\noindent {\bf Reconstruction and simulation of neocortical microcircuitry} \cite{Markram2015}: This model is a reconstruction of the microcircuitry of the somatosensory cortex of juvenile rat. It is based on an algorithm developed by the authors to reconstruct detailed anatomy and physiology from sparse experimental data \cite{Reimann2015}. The model represents a neocortical volume of 0.29 $\pm$ 0.01 mm$^3$ containing $\sim$31,000 neurons, with 55 layer-specific morphological and 207 morphoelectrical neuron subtypes distributed in the 6 cortical layers. Neurons are described by multicompartmental conductance-based models using up to 13 active ion channel types and an intracellular Ca$^{2+}$ dynamics. The network connections were created by positioning the neurons in the volume and setting the synaptic contacts among the overlapping arbors according to biological constrains, which resulted in about 37 million synapses. Excitatory synapses are modeled using both AMPA and NMDA receptor kinetics, and inhitibory synapses using a combination of GABA$_\text{A}$ and GABA$_\text{B}$ receptor kinetics. The model reproduces several \emph{in vitro} and \emph{in vivo} experiments without parameter tuning. Simulations also show transition from synchronous to asynchronous activity modulated by physiological mechanisms. The reconstructed model and experimental data are available at: https://bbp.epfl.ch/nmc-portal \cite{Ramaswamy2015}.
\newline
    
\noindent {\bf Data-driven modeling of cholinergic modulation of neural microcircuits: bridging neurons, synapses and network activity} \cite{Ramaswamy2018}: 
This model incorporates cholinergic modulation to the microcirtuit model described above \cite{Markram2015,Reimann2015,Ramaswamy2015}. The modulation by acetylcholine (ACh) on cellular excitability is implemented by a depolarizing step current injection and, on synaptic transmission, by changing the utilization of synaptic efficacy parameter in the used stochastic synapse model. In the untuned version of model, Ach desynchronizes spontaneous network activity. Simulations show that, at low level of Ach, the network exhibits slow oscillations and network synchrony as observed in non-rapid eye movement (nREM) sleep, and, at high ACh concentrations, it exhibits fast oscillations and network asynchrony as during wakefulness and REM sleep. Data analysis as the cross-correlograms were computed using MATLAB (The MathWorks Inc., Natick, USA).
\newline

\noindent {\bf Virtual Electrode Recording Tool for EXtracellular potentials (VERTEX): comparing multi-electrode recordings from simulated and biological mammalian cortical tissue} \cite{Tomsett2015}: This model contains $175,421$ excitatory and inhibitory neurons subdivided in 15 different cell types distributed in five layers (2/3, 4, 5 and 6) with several characteristics based on cat V1 data \cite{Binzegger2004,Izhikevich2008}. The neurons are described by reduced compartmental models \cite{Mainen1996,Bush1993}. The neuronal somata are randomly placed in a 3D space according to a rule based on the cortical depth. Synaptic connections are created based on distance, following a Gaussian kernel distribution with a fixed radius, and the number of synapses between groups of neurons is adapted from experimental data \cite{Binzegger2004}. The synapses are defined as AMPA for excitatory and GABA for inhibitory connections, and they are modeled as conductance-based with exponential decay. The authors developed a tool implemented in MATLAB (The MathWorks Inc., Natick, USA) to obtain the LFP from the model, the Virtual Electrode Recording Tool for EXtracellular potentials (VERTEX), which preserves the spatial and frequency-scaling features of the LFP. Codes are available in the resources section at \url{https://www.dynamic-connectome.org/}.
\newline
    
\noindent {\bf Impact of higher order network structure on emergent cortical activity} \cite{Nolte2019}:
This model is based on the detailed neocortical microcircuit model of Markram \emph{et al.} described above \cite{Markram2015}. From this detailed model, the authors generated another model with similar first-order structure but unconstrained higher order characteristics. This resulted in two models with different synaptic connectivity. The model by Markram \emph{et al.} includes higher order structures, e.g. abundance of cliques of all-to-all connected neurons, arising from the morphological diversity of neuronal types, and the model generated by the authors does not have such higher order structures. By analyzing the two models, the authors demonstrated that the differences in higher order network structure have an impact on emergent cortical activity, and that the presence of higher order structure leads to more meaningful neuronal firing patters such as increased pairwise correlation. The main conclusion is that the higher order structure created by morphological diversity within neuronal types has an impact on emergent cortical activity.

\section{Conclusion}

In this tutorial review, we covered details about how to build large-scale neuronal network models with special emphasis on the connectivity maps. Recent models for which connectivity is an essential aspect were reviewed and commented. Modeling neurons, synapses, and networks is a problem of increasing complexity, which is mostly derived from neurobiological recordings \cite{BasGaz11}. Among the different choices a modeler has to make, there are several guides helping these decisions \cite{DayAbb2001,BreRud07,blohm2020}. Our focus, however, is on networks composed of spiking-neurons where the connectivity is determined from a map that is based on experiments (a data-driven connectivity map). To the best of our knowledge, there is no review exploring this important step in a tutorial manner.

Before addressing the connectivity, we discussed concisely how neuron and synaptic models can be tackled. Depending on the aimed level of biological detail, neurons can be modeled containing ionic channels or not. The same strategy can be forwarded to neurons with or without morphology. Moving to synapses, the choice of synaptic model and whether it includes plasticity or not adds another dimension of complexity to the problem. Synaptic models may introduce time-dependent properties such as short-term plasticity or even spike-timing-dependent plasticity where the difference between pre- and postsynaptic spike-timing determines strengthening or weakening of a synapse. Regardless of the single neuron or synaptic properties, every neuron and synapse model can be used in the construction of a network, which, depending on the modeling compromises assumed, results in a enormous repertoire of possible degrees of model complexity.

Proceeding from neurons and synapses to the network connectivity, the researcher in possession of the experimental data, i.e. the connectivity map, will need to interpret it and translate its information into synaptic links among neurons. As we have shown in the comparison between results obtained with Eqs.~\ref{Eq:conn} and~\ref{Eq:conn_approx}, care must be taken when using approximate expressions for the connection probabilities. The differences between networks generated with the exact and the approximate connection probability expressions may go unnoticed when the network size is small, but as the network size grows they become evident and may have great impact of the neuronal activity. 

Similarly, the role of the spatial organization or of the morphological organization may be critical in determining how neurons are connectivity. First, placing neurons in a spatially structured manner implies that neurons are more likely to be connected to closer neighbours than to the farther ones. In light of that, several works approach the problem by drawing connections with distant-dependent connection probabilities that decay with distance. Secondly, the neuronal structure is highly important especially when considered together with the electrophysiological class of the neuron. For instance, basket cells in the cortex make synaptic contacts with the somata of pyramidal cells whereas bitufted and Martinotti cells usually make their synaptic contacts to pyramidal cells in the dendrites of these cells \cite{HofJac17}. Hence specificity is an essential aspect of connectivity maps.

It is important to emphasize the restrictions imposed by the brain anatomy to the network models (e.g. geometric constraints due to the limited size of the cranium and finite size of connections, differential effects of distal vs. proximal dendrites, etc). Although it is usual and reasonable to make approximations to deal with these anatomical features when building a model, frequently they are ignored and not even mentioned omitting relevant information that may be important to neuroscience. Modellers must be aware of the possible consequences of not taking into consideration anatomical elements and should consider them when making conclusions from their network models.  

Regarding the tool(s) used to implement the model, a modeler either approaches the problem with a low-level language such as C/C++ or uses a neurosimulator which encompasses high-level functions that are easy to interpret and favor code sharing with a larger community. Among the multitude of available neurosimulators, we discussed here three examples which are currently popular: Brian 2, NEST, and NEURON. Searches made on \emph{Google Scholar} for the keywords ``NEURON simulator'', ``NEST simulator'' and ``Brian 2 simulator'' within the period 2015-2020 gave, respectively, 409, 377 and 35 results. For comparison, a similar search for ``spiking network model'' resulted in 812 hits. All searches were made on December 2nd, 2020. Our discussion of these neurosimulators highlighted their advantages and disadvantages with respect to specific modeling situations.

Finally, we gave at the end a list of cutting-edge recent works in which the data-driven approach to construct large-scale networks of spiking neurons was used. The list contains brief summaries of the characteristics of the models and how they were used. We hope the information given in Section~\ref{sect:modeling} is sufficient to help the reader, especially a newcomer to the field of computational neuroscience, to start reading and understanding these works. We also hope our information is useful to guide the choice and use of one of the publicly available models in a research project. 

\bigskip

\section*{Acknowledgements}

This article was produced as part of the S. Paulo Research Foundation (FAPESP) Research, Innovation and Dissemination Center for Neuromathematics (CEPID NeuroMat, Grant No. 2013/07699-0). 
The authors also thank FAPESP support through Grants No. 
2013/25667-8 (R.F.O.P.),
2015/50122-0 and 2018/20277-0 (A.C.R.),
2016/03855-5 (N.L.K.), 
2017/07688-9 (R.O.S),
and
2018/09150-9 (M.G.-S.).
V.L. is supported by a CAPES PhD scholarship.
A.C.R. thanks financial support from the National Council of Scientific and Technological Development (CNPq), Grant No. 306251/2014-0.
This study was financed in part by the Coordenação de Aperfeiçoamento de Pessoal de Nível Superior - Brasil (CAPES) - Finance Code 001.

\section*{Author contribution statement}

Author Contributions: R.O.S., R.F.O.P., V.L., N.L.K., M.G.-S. and A.C.R.: Conceived the work; R.O.S., R.F.O.P., V.L., N.L.K., M.G.-S. and A.C.R.: literature search; R.O.S., R.F.O.P., V.L., N.L.K., M.G.-S. and A.C.R.: Wrote the manuscript. All authors read and agreed to the published version of the manuscript.

\bibliographystyle{epj}
%\bibliography{ConnectivityReview}

\begin{thebibliography}{193}

\bibitem{Azevedo2009}
F.A.C. Azevedo, L.R.B. Carvalho, L.T. Grinberg, J.M. Farfel, R.E.L. Ferretti,
  R.E.P. Leite, W.~Jacob~Filho, R.~Lent, S.~Herculano-Houzel, J. Comp. Neurol.
  \textbf{513(5)}, 532 (2009)

\bibitem{BraSchuz1998}
V.~Braitenberg, A.~Schüz, \emph{Cortex: {S}tatistics and {G}eometry of
  {N}euronal {C}onnectivity} (Springer, Berlin, 1998)

\bibitem{Tononi1994}
G.~Tononi, O.~Sporns, G.M. Edelman, P. Natl. Acad. Sci. USA \textbf{91}(11),
  5033 (1994), ISSN 00278424

\bibitem{Tononi1998}
G.~Tononi, G.M. Edelman, O.~Sporns, \emph{{Complexity and coherency:
  Integrating information in the brain}} (1998)

\bibitem{Koch1999}
C.~Koch, G.~Laurent, \emph{{Complexity and the nervous system}} (1999)

\bibitem{BasGaz11}
D.S. Bassett, M.S. Gazzaniga, Trends Cogn. Sci. \textbf{15}(5), 200 (2011)

\bibitem{barkemaMC}
M.E.J. Newman, G.T. Barkema, \emph{{M}onte {C}arlo Methods in Statistical
  Physics} (Oxford University Press, New York, UK, 1999)

\bibitem{gennesLC1993}
P.G. de{ }Gennes, J.~Prost, \emph{The {P}hysics of the {L}iquid {C}rystals}
  (Oxford University Press, Oxford, UK, 1993)

\bibitem{odorReview2004}
G.~{\'O}dor, Rev. Mod. Phys. \textbf{76(3)}, 663 (2004)

\bibitem{mallinson2019}
J.B. Mallinson, S.~Shirai, S.K. Acharya, S.K. Bose, E.~Galli, S.A. Brown, Sci.
  Adv. \textbf{5}, eaaw8438 (2019)

\bibitem{Johnston1995}
D.~Johnston, S.M.S. Wu, \emph{{F}oundations of {C}ellular {N}europhysiology}
  (The MIT Press, Cambridge, MA, 1995)

\bibitem{sporns2010}
O.~Sporns, \emph{Networks of the Brain} (MIT press, Cambridge, MA, 2010)

\bibitem{TomPen14}
P.~Tomov, R.F. Pena, M.A. Zaks, A.C. Roque, Front. Comput. Neurosci.
  \textbf{8}, 103 (2014)

\bibitem{GirardiPLR2020}
M.~Girardi{-}Schappo, A.~de~Andrade~Costa, Phys. Life Rev. \textbf{33}, 19
  (2020)

\bibitem{Hesse2014}
J.~Hesse, T.~Gross, Front. Syst. Neurosci. \textbf{8}, 166 (2014)

\bibitem{Carvalho2021}
T.T.A. Carvalho, A.J. Fontenele, M.~Girardi-Schappo, T.~Feliciano, L.A.A.
  Aguiar, T.P.L. Silva, N.A.P. de~Vasconcelos, P.V. Carelli, M.~Copelli, Front.
  Neural Circuits \textbf{14}, 576727 (2021)

\bibitem{Tibau2013}
E.~Tibau, C.~Bendiksen, S.~Teller, N.~Amigó, J.~Soriano, AIP Conf. Proc.
  \textbf{1510}(1), 54 (2013)

\bibitem{yamamoto2018}
H.~Yamamoto, S.~Moriya, K.~Ide, T.~Hayakawa, H.~Akima, S.~Sato, S.~Kubota,
  T.~Tanii, M.~Niwano, S.~Teller, J.~Soriano, A.~Hirano{-}Iwata, Sci. Adv.
  \textbf{4}, eaau4914 (2018)

\bibitem{girardiV1conf2015}
G.S. Bortolotto, M.~Girardi{-}Schappo, J.J. Gonsalves, L.T. Pinto, M.H.R.
  Tragtenberg, J. Phys. Conf. Ser. \textbf{686}(1), 012008 (2016)

\bibitem{muller2018cortical}
L.~Muller, F.~Chavane, J.~Reynolds, T.J. Sejnowski, Nat. Rev. Neurosci.
  \textbf{19}(5), 255 (2018)

\bibitem{Girardi2016}
M.~Girardi{-}Schappo, G.S. Bortolotto, J.J. Gonsalves, L.T. Pinto, M.H.R.
  Tragtenberg, Sci. Rep. \textbf{6}, 29561 (2016)

\bibitem{arnulfo2020long}
G.~Arnulfo, S.H. Wang, V.~Myrov, B.~Toselli, J.~Hirvonen, M.M. Fato, L.~Nobili,
  F.~Cardinale, A.~Rubino, A.~Zhigalov, S.~Palva, J.M. Palva, Nat. Commun.
  \textbf{11}(1), 1 (2020)

\bibitem{Girardi2018}
M.~Girardi{-}Schappo, M.H.R. Tragtenberg, Phys. Rev. E \textbf{97}, 042415
  (2018)

\bibitem{lameu2012}
E.L. Lameu, C.~Batista, A.~Batista, K.~Iarosz, R.~Viana, S.~Lopes, J.~Kurths,
  Chaos \textbf{22}, 043149 (2012)

\bibitem{Kaiser2007}
M.~Kaiser, M.~G{\"o}rner, C.C. Hilgetag, New J. Phys. \textbf{9}(5), 110 (2007)

\bibitem{pena2020}
R.F. Pena, V.~Lima, R.O. Shimoura, J.P. Novato, A.C. Roque, Brain Sci.
  \textbf{10}(4), 228 (2020)

\bibitem{lameu2019}
E.L. Lameu, F.S. Borges, K.C. Iarosz, P.R. Protachevicz, A.M. Batista, C.G.
  Antonopoulos, E.E.N. Macau, arXiv pp. 1911.00052 [q--bio.NC] (2019)

\bibitem{rabinowitch2019would}
I.~Rabinowitch, Phys. Life Rev.  (2019)

\bibitem{Roxin2004}
A.~Roxin, H.~Riecke, S.A. Solla, Phys. Rev. Lett \textbf{92}(19), 198101 (2004)

\bibitem{Lin2005}
M.~Lin, T.~Chen, Phys. Rev. E \textbf{71}, 016133 (2005)

\bibitem{BorPro17}
F.~Borges, P.~Protachevicz, E.~Lameu, R.~Bonetti, K.~Iarosz, I.~Caldas,
  M.~Baptista, A.~Batista, Neural Netw. \textbf{90}, 1 (2017)

\bibitem{Bezaire2016}
M.J. Bezaire, I.~Raikov, K.~Burk, D.~Vyas, I.~Soltesz, eLife \textbf{5}, e18566
  (2016)

\bibitem{DuraBernal2017}
S.~Dura-Bernal, S.~Neymotin, B.~Suter, G.~Shepherd, W.~Lytton, bioRxiv p.
  201707 (2017)

\bibitem{Brunton2019}
B.W. Brunton, M.~Beyeler, Curr. Opin. Neurobiol. \textbf{58}, 21 (2019)

\bibitem{Segev1998}
I.~Segev, R.E. Burke, in \emph{Methods in {N}euronal {M}odeling: From Ions to
  Networks, 2nd. Edition}, edited by C.~Koch, I.~Segev (The MIT Press Press,
  Cambridge, MA, 1998), pp. 93--136

\bibitem{Herz2006}
A.V.M. Herz, T.~Gollisch, C.K. Machens, D.~Jaeger, Science \textbf{314}, 80
  (2006)

\bibitem{sterratt2012}
D.~Sterratt, B.~Graham, A.~Gillies, D.~Willshaw, \emph{Principles of
  {C}omputational {M}odelling in {N}euroscience} (Cambridge University Press,
  2012)

\bibitem{brunel2000}
N.~Brunel, J. Comput. Neurosci. \textbf{8}, 183 (2000)

\bibitem{gerstner2014}
W.~Gerstner, W.M. Kistler, R.~Naud, L.~Paninski, \emph{Neuronal {D}ynamics:
  From Single Neurons to Networks and Models of Cognition} (Cambridge
  University Press, 2014)

\bibitem{Potjans2014}
T.C. Potjans, M.~Diesmann, Cereb. Cortex \textbf{24}, 785 (2014)

\bibitem{DayAbb2001}
P.~Dayan, L.F. Abbott, \emph{Theoretical Neuroscience: Computational and
  Mathematical Modeling of Neural Systems} (The MIT Press, Cambridge,
  Massachussetts, USA, 2001)

\bibitem{Liley2015}
D.T.J. Liley, in \emph{Encyclopedia of {C}omputational {N}euroscience}, edited
  by D.~Jaeger, R.~Jung (Springer, New York, NY, 2015), pp. 1898--1912

\bibitem{Cowan2016}
J.D. Cowan, J.~Neuman, W.~van Drongelen, J. Math. Neurosc. \textbf{6}, 1 (2016)

\bibitem{ostojic2014}
S.~Ostojic, Nat. Neurosci. \textbf{17}, 594 (2014)

\bibitem{pena2018}
R.F. Pena, M.A. Zaks, A.C. Roque, J. Comput. Neurosci. \textbf{45}(1), 1 (2018)

\bibitem{BorPro20}
F.S. Borges, P.R. Protachevicz, R.F. Pena, E.L. Lameu, G.S. Higa, A.H. Kihara,
  F.S. Matias, C.G. Antonopoulos, R.~de~Pasquale, A.C. Roque, K.C. Iarosz,
  P.~Ji, A.M. Batista, Physica A \textbf{537}, 122671 (2020)

\bibitem{destexhe2001}
A.~Destexhe, M.~Rudolph, J.M. Fellous, T.J. Sejnowski, Neuroscience
  \textbf{107}, 13 (2001)

\bibitem{Vogels2005}
T.P. Vogels, L.F. Abbott, J. Neurosci. \textbf{25}, 10786 (2005)

\bibitem{buzsaki2010neural}
G.~Buzs{\'a}ki, Neuron \textbf{68}(3), 362 (2010)

\bibitem{papadimitriou2020brain}
C.H. Papadimitriou, S.S. Vempala, D.~Mitropolsky, M.~Collins, W.~Maass, P.
  Natl. Acad. Sci. USA  (2020)

\bibitem{sporns2005}
O.~Sporns, G.~Tononi, R.~K{\"o}tter, PLoS Comput. Biol. \textbf{1}(4), e42
  (2005)

\bibitem{Bullmore2011}
E.T. Bullmore, D.S. Bassett, Annu. Rev. Clin. Psycho. \textbf{7}, 113 (2011)

\bibitem{watts1998collective}
D.J. Watts, S.H. Strogatz, Nature \textbf{393}(6684), 440 (1998)

\bibitem{mountcastle1997}
V.B. Mountcastle, Brain \textbf{120}(4), 701 (1997)

\bibitem{kaiser2010}
M.~Kaiser, C.C. Hilgetag, Front. Neuroinform. \textbf{4}, 8 (2010)

\bibitem{meunier2010}
D.~Meunier, R.~Lambiotte, E.T. Bullmore, Front. Neurosci. \textbf{4} (2010)

\bibitem{TomPen16}
P.~Tomov, R.F. Pena, A.C. Roque, M.A. Zaks, Front. Comput. Neurosci.
  \textbf{10}, 23 (2016)

\bibitem{Shepherd2018}
G.M. Shepherd, S.~Grillner, eds., \emph{Handbook of Brain Microcircuits, Second
  edition} (Oxford University Press, New York, NY, USA, 2018)

\bibitem{ThoWes02}
A.M. Thomson, D.C. West, Y.~Wang, A.P. Bannister, Cereb. Cortex \textbf{12}(9),
  936 (2002)

\bibitem{Binzegger2004}
T.~Binzegger, R.J. Douglas, K.A.C. Martin, J. Neurosci. \textbf{24}, 8441
  (2004)

\bibitem{Li2013}
Y.T. Li, M.~Zhou, H.W. Tao, L.I. Zhang, Nat. Neurosci. \textbf{16}, 1179 (2013)

\bibitem{lien2013}
A.D. Lien, M.~Scanziani, Nat. Neurosci. \textbf{16}, 1315 (2013)

\bibitem{sporns2011human}
O.~Sporns, Ann. NY Acad. Sci. \textbf{1224}(1), 109 (2011)

\bibitem{betzel2020network}
R.~Betzel, arXiv preprint arXiv:2010.01591  (2020)

\bibitem{PaxHua00}
G.~Paxinos, X.F. Huang, A.W. Toga, \emph{The {R}hesus {M}onkey {B}rain in
  {S}tereotaxic {C}oordinates} (Academic Press, San Diego, CA, 2000)

\bibitem{AliChu13}
A.P. Alivisatos, M.~Chun, G.M. Church, K.~Deisseroth, J.P. Donoghue, R.J.
  Greenspan, P.L. McEuen, M.L. Roukes, T.J. Sejnowski, P.S. Weiss, R.~Yuste,
  Science \textbf{339}(6125), 1284 (2013)

\bibitem{Ste13}
K.E. Stephan, NeuroImage \textbf{80}, 46 (2013)

\bibitem{Basser1994}
P.J. Basser, J.~Mattiello, D.~LeBihan, Biophys. J. \textbf{66}(1), 259 (1994)

\bibitem{Markram2015}
H.~Markram, E.~Muller, S.~Ramaswamy, M.W. Reimann, M.~Abdellah, C.A. Sanchez,
  A.~Ailamaki, L.~Alonso-Nanclares, N.~Antille, S.~Arsever, G.A.A. Kahou, T.K.
  Berger, A.~Bilgili, N.~Buncic, A.~Chalimourda, G.~Chindemi, J.D.D. Courcol,
  F.~Delalondre, V.~Delattre, S.~Druckmann, R.~Dumusc, J.~Dynes, S.~Eilemann,
  E.~Gal, M.E. Gevaert, J.P.P. Ghobril, A.~Gidon, J.W. Graham, A.~Gupta,
  V.~Haenel, E.~Hay, T.~Heinis, J.B. Hernando, M.~Hines, L.~Kanari, D.~Keller,
  J.~Kenyon, G.~Khazen, Y.~Kim, J.G. King, Z.~Kisvarday, P.~Kumbhar,
  S.~Lasserre, J.V. {Le B{\'{e}}}, B.R. Magalh{\~{a}}es,
  A.~Merch{\'{a}}n-P{\'{e}}rez, J.~Meystre, B.R. Morrice, J.~Muller,
  A.~Mu{\~{n}}oz-C{\'{e}}spedes, S.~Muralidhar, K.~Muthurasa, D.~Nachbaur, T.H.
  Newton, M.~Nolte, A.~Ovcharenko, J.~Palacios, L.~Pastor, R.~Perin, R.~Ranjan,
  I.~Riachi, J.R.R. Rodr{\'{i}}guez, J.L. Riquelme, C.~R{\"{o}}ssert,
  K.~Sfyrakis, Y.~Shi, J.C. Shillcock, G.~Silberberg, R.~Silva, F.~Tauheed,
  M.~Telefont, M.~Toledo-Rodriguez, T.~Tr{\"{a}}nkler, W.~{Van Geit}, J.V.
  D{\'{i}}az, R.~Walker, Y.~Wang, S.M. Zaninetta, J.~Defelipe, S.L. Hill,
  I.~Segev, F.~Sch{\"{u}}rmann, J.V. {Le B{\'{e}}}, B.R. Magalh{\~{a}}es,
  A.~Merch{\'{a}}n-P{\'{e}}rez, J.~Meystre, B.R. Morrice, J.~Muller,
  A.~Mu{\~{n}}oz-C{\'{e}}spedes, S.~Muralidhar, K.~Muthurasa, D.~Nachbaur, T.H.
  Newton, M.~Nolte, A.~Ovcharenko, J.~Palacios, L.~Pastor, R.~Perin, R.~Ranjan,
  I.~Riachi, J.R.R. Rodr{\'{i}}guez, J.L. Riquelme, C.~R{\"{o}}ssert,
  K.~Sfyrakis, Y.~Shi, J.C. Shillcock, G.~Silberberg, R.~Silva, F.~Tauheed,
  M.~Telefont, M.~Toledo-Rodriguez, T.~Tr{\"{a}}nkler, W.~{Van Geit}, J.V.
  D{\'{i}}az, R.~Walker, Y.~Wang, S.M. Zaninetta, J.~Defelipe, S.L. Hill,
  I.~Segev, F.~Sch{\"{u}}rmann, Cell \textbf{163}, 456 (2015)

\bibitem{Kuypers1990}
H.~Kuypers, G.~Ugolini, Trends Neurosci. \textbf{13}(2), 71 (1990)

\bibitem{Saleeba2019}
C.~Saleeba, B.~Dempsey, S.~Le, A.~Goodchild, S.~McMullan, Front. Neurosci.
  \textbf{13}, 897 (2019)

\bibitem{Denk2004}
W.~Denk, H.~Horstmann, PLoS Biol. \textbf{2}(11) (2004)

\bibitem{SheHar20}
F.Y. Shen, M.M. Harrington, L.A. Walker, H.P.J. Cheng, E.S. Boyden, D.~Cai,
  Nat. Commun. \textbf{11}, 4632 (2020)

\bibitem{VanAlbada2020}
S.J. van Albada, A.~Morales-Gregorio, T.~Dickscheid, A.~Goulas, R.~Bakker,
  S.~Bludau, G.~Palm, C.C. Hilgetag, M.~Diesmann, arXiv p. 2007.00031 (2020)

\bibitem{andersen2006}
P.~Andersen, R.~Morris, D.~Amaral, T.~Bliss, J.~O'Keefe, \emph{The
  {H}ippocampus {B}ook} (Oxford University Press, 2006)

\bibitem{kajiwara2008}
R.~Kajiwara, F.G. Wouterlood, A.~Sah, A.J. Boekel, L.T. Baks-te Bulte, M.P.
  Witter, Hippocampus \textbf{18}(3), 266 (2008)

\bibitem{van2009}
N.~Van~Strien, N.~Cappaert, M.~Witter, Nat. Rev. Neurosci. \textbf{10}(4), 272
  (2009)

\bibitem{Witter2010}
M.P. Witter, in \emph{Hippocampal Microcircuits} (Springer, 2010), pp. 5--26

\bibitem{maller2019}
J.J. Maller, T.~Welton, M.~Middione, F.M. Callaghan, J.V. Rosenfeld, S.M.
  Grieve, Sci. Rep. \textbf{9}(1), 1 (2019)

\bibitem{Amunts2019}
K.~Amunts, A.C. Knoll, T.~Lippert, C.M.A. Pennartz, P.~Ryvlin, A.~Destexhe,
  V.K. Jirsa, E.~D'Angelo, J.G. Bjaalie, PLoS Biol. \textbf{17}(e3000344)
  (2019)

\bibitem{Markram2006}
H.~Markram, Nat. Rev. Neurosci. \textbf{7}(153–160) (2006)

\bibitem{Wang2020}
Q.~Wang, S.L. Ding, Y.~Li, J.~Royall, D.~Feng, P.~Lesnar, N.~Graddis,
  M.~Naeemi, B.~Facer, A.~Ho, T.~Dolbeare, B.~Blanchard, N.~Dee, W.~Wakeman,
  K.E. Hirokawa, A.~Szafer, S.M. Sunkin, S.W. Oh, A.~Bernard, J.W. Phillips,
  M.~Hawrylycz, C.~Koch, H.~Zeng, J.A. Harris, L.~Ng, Cell \textbf{181}, 936
  (2020)

\bibitem{kandel2013neuroscience}
E.R. Kandel, H.~Markram, P.M. Matthews, R.~Yuste, C.~Koch, Nat. Rev. Neurosci.
  \textbf{14}(9), 659 (2013)

\bibitem{landhuis2017neuroscience}
E.~Landhuis, Nature \textbf{541}, 559–561 (2017)

\bibitem{plesser2007efficient}
H.E. Plesser, J.M. Eppler, A.~Morrison, M.~Diesmann, M.O. Gewaltig,
  \emph{Efficient Parallel Simulation of Large-scale Neuronal Networks on
  Clusters of Multiprocessor Computers}, in \emph{European Conference on
  Parallel Processing} (Springer, 2007), pp. 672--681

\bibitem{kunkel2014spiking}
S.~Kunkel, M.~Schmidt, J.M. Eppler, H.E. Plesser, G.~Masumoto, J.~Igarashi,
  S.~Ishii, T.~Fukai, A.~Morrison, M.~Diesmann, M.~Helias, Front. Neuroinform.
  \textbf{8}, 78 (2014)

\bibitem{schmidt2015full}
M.~Schmidt, R.~Bakker, K.~Shen, G.~Bezgin, C.C. Hilgetag, M.~Diesmann, S.J. van
  Albada, arXiv preprint arXiv:1511.09364  (2015)

\bibitem{She2017neo}
G.M. Shepherd, T.B. Rowe, Front. Neuroanat. \textbf{11}, 100 (2017)

\bibitem{Izhikevich2003}
E.M. Izhikevich, IEEE Trans. Neural Netw. \textbf{14}, 1569 (2003)

\bibitem{Brette2005}
R.~Brette, W.~Gerstner, J. Neurophysiol. \textbf{94}, 3637 (2005)

\bibitem{pena2018b}
R.F. Pena, C.C. Ceballos, V.~Lima, A.C. Roque, Phys. Rev. E \textbf{97}(4),
  042408 (2018)

\bibitem{Girardi2013}
M.~Girardi-Schappo, M.H.R. Tragtenberg, O.~Kinouchi, J. Neurosci. Meth.
  \textbf{220}, 116 (2013)

\bibitem{Girardi2017}
M.~Girardi-Schappo, G.S. Bortolotto, R.V. Stenzinger, J.J. Gonsalves, M.H.R.
  Tragtenberg, PLoS ONE \textbf{12}, e0174621 (2017)

\bibitem{HodHux52}
A.L. Hodgkin, A.F. Huxley, J. Physiol. \textbf{117}(4), 500 (1952)

\bibitem{Roth2010}
A.~Roth, C.W. van Rossum, in \emph{Computational {M}odeling {M}ethods for
  {N}euroscientists}, edited by E.~De~Schutter (The MIT Press, Cambridge, MA,
  2010), pp. 139--159

\bibitem{TsoPaw98}
M.~Tsodyks, K.~Pawelzik, H.~Markram, Neural Comput. \textbf{10}, 821 (1998)

\bibitem{castellani2001}
G.C. Castellani, E.M. Quinlan, L.N. Cooper, H.Z. Shouval, Proc. Natl. Acad.
  Sci. USA \textbf{98}, 12772 (2001)

\bibitem{tsodyks2005}
M.~Tsodyks, in \emph{Methods and {M}odels in {N}europhysics}, edited by
  C.~Chow, B.~Gutkin, D.~Hansel, C.~Meunier, J.~Dalibard (Elsevier, 2005),
  Vol.~80 of \emph{Les Houches}, pp. 245 -- 265,
  \texttt{http://www.sciencedirect.com/science/article/pii/S0924809905800137}

\bibitem{clopath2010}
C.~Clopath, L.~Busing, E.~Vasilaki, W.~Gerstner, Nat. Neurosci. \textbf{13},
  344 (2010)

\bibitem{MarLub97}
H.~Markram, J.~L{\"u}bke, M.~Frotscher, B.~Sakmann, Science \textbf{275}(5297),
  213 (1997)

\bibitem{MorDie08}
A.~Morrison, M.~Diesmann, W.~Gerstner, Biol. Cybern. \textbf{98}(6), 459 (2008)

\bibitem{sjo10}
J.~Sj{\"o}str{\"o}m, W.~Gerstner, Scholarpedia \textbf{5}, 1362 (2010)

\bibitem{SjoPer01}
P.J. Sj{\"o}str{\"o}m, G.G. Turrigiano, S.B. Nelson, Neuron \textbf{32}(6),
  1149 (2001)

\bibitem{shimoura2015}
R.O. Shimoura, R.F. Pena, A.C. Roque, BMC Neurosci. \textbf{16}(S1), P210
  (2015)

\bibitem{KepVan02}
A.~Kepecs, M.C. van Rossum, S.~Song, J.~Tegner, Biol. Cybern. \textbf{87}(5-6),
  446 (2002)

\bibitem{KleFuk14}
F.I. Kleberg, T.~Fukai, M.~Gilson, Front. Comput. Neurosci. \textbf{8}, 53
  (2014)

\bibitem{Schmidt2018a}
M.~Schmidt, R.~Bakker, K.~Shen, G.~Bezgin, M.~Diesmann, S.J. van Albada, PLoS
  Comput. Biol. \textbf{14}, e1006359 (2018)

\bibitem{Schmidt2018b}
M.~Schmidt, R.~Bakker, C.C. Hilgetag, M.~Diesmann, S.J. van Albada, Brain
  Struct. Funct. \textbf{223}, 1409 (2018)

\bibitem{Newman2010Book}
M.E.J. Newman, \emph{Networks: An Introduction} (Oxford University Press, New
  York, UK, 2010)

\bibitem{Shimoura2018}
R.O. Shimoura, N.L. Kamiji, R.F. Pena, V.L. Cordeiro, C.C. Ceballos,
  R.~Cecilia, A.C. Roque, ReScience \textbf{4} (2018)

\bibitem{Billeh2020}
Y.N. Billeh, B.~Cai, S.L. Gratiy, K.~Dai, R.~Iyer, N.W. Gouwens, R.~Abbasi-Asl,
  X.~Jia, J.H. Siegle, S.R. Olsen, C.~Koch, S.~Mihalas, A.~Arkhipov, Neuron
  \textbf{106}, 388 (2020)

\bibitem{kawaguchi1997}
Y.~Kawaguchi, Y.~Kubota, Cereb. Cortex \textbf{7}(6), 476 (1997)

\bibitem{tremblay2016}
R.~Tremblay, S.~Lee, B.~Rudy, Neuron \textbf{91}(2), 260 (2016)

\bibitem{safari2017}
M.S. Safari, J.~Mirnajafi-Zadeh, H.~Hioki, T.~Tsumoto, Sci. Rep. \textbf{7}(1),
  1 (2017)

\bibitem{Leao2012}
R.N. Leão, S.~Mikulovic, K.E. Leão, H.~Munguba, H.~Gezelius, A.~Enjin,
  K.~Patra, A.~Eriksson, L.M. Loew, A.B. Tort, K.~Kullander, Nat. Neurosci.
  \textbf{15}, 1524– (2012)

\bibitem{glaser1990neuron}
J.R. Glaser, E.M. Glaser, Comput. Med. Imag. Grap. \textbf{14}(5), 307 (1990)

\bibitem{stockley1993}
E.~Stockley, H.~Cole, A.~Brown, H.~Wheal, J. Neurosci. Meth. \textbf{47}(1-2),
  39 (1993)

\bibitem{crook2007morphml}
S.~Crook, P.~Gleeson, F.~Howell, J.~Svitak, R.A. Silver, Neuroinformatics
  \textbf{5}(2), 96 (2007)

\bibitem{Hines1997}
M.L. Hines, N.T. Carnevale, Neural Comput. \textbf{9}, 1179 (1997)

\bibitem{gleeson2007neuroconstruct}
P.~Gleeson, V.~Steuber, R.A. Silver, Neuron \textbf{54}(2), 219 (2007)

\bibitem{bower2012book}
J.M. Bower, D.~Beeman, \emph{The Book of GENESIS: Exploring Realistic Neural
  Models with the GEneral NEural SImulation System} (Springer Science \&
  Business Media, 2012)

\bibitem{Crone2019}
J.C. Crone, M.M. Vindiola, A.B. Yu, D.L. Boothe, D.~Beeman, K.S. Oie, P.J.
  Franaszczuk, Front. Neuroinform. \textbf{13}, 15 (2019)

\bibitem{Bhalla2008}
U.S. Bhalla, Front. Neuroinform. \textbf{2}, 6 (2008)

\bibitem{Ray2008}
S.~Ray, R.~Deshpande, N.~Dudani, U.S. Bhalla, BMC Neurosci. \textbf{9}(Suppl
  1), P93 (2008)

\bibitem{cannon2010sto}
R.C. Cannon, C.~O'Donnell, M.F. Nolan, PLoS Comput. Biol. \textbf{6}, e1000886
  (2010)

\bibitem{davison2009pynn}
A.P. Davison, D.~Brüderle, J.~Eppler, J.~Kremkow, E.~Muller, D.~Pecevski,
  L.~Perrinet, P.~Yger, Front. Neuroinf. \textbf{2}, 11 (2009)

\bibitem{heitmann2018brain}
S.~Heitmann, M.J. Aburn, M.~Breakspear, Neurocomput. \textbf{315}, 82 (2018)

\bibitem{Nordlie2009}
E.~Nordlie, M.O. Gewaltig, H.E. Plesser, PLoS Comput. Biol. \textbf{5},
  e1000456 (2009)

\bibitem{McDougal2016}
R.A. McDougal, A.S. Bulanova, W.W. Lytton, IEEE Trans. Biomed. Eng.
  \textbf{63}, 2021–2035 (2016)

\bibitem{robitaille2013astropy}
T.P. Robitaille, E.J. Tollerud, P.~Greenfield, M.~Droettboom, E.~Bray,
  T.~Aldcroft, M.~Davis, A.~Ginsburg, A.M. Price-Whelan, W.E. Kerzendorf,
  A.~Conley, N.~Crighton, K.~Barbary, D.~Muna, H.~Ferguson, F.~Grollier, M.M.
  Parikh, P.H. Nair, H.M. G{\"{u}}nther, C.~Deil, J.~Woillez, S.~Conseil,
  R.~Kramer, J.E. Turner, L.~Singer, R.~Fox, B.A. Weaver, V.~Zabalza, Z.I.
  Edwards, K.~{Azalee Bostroem}, D.J. Burke, A.R. Casey, S.M. Crawford,
  N.~Dencheva, J.~Ely, T.~Jenness, K.~Labrie, P.L. Lim, F.~Pierfederici,
  A.~Pontzen, A.~Ptak, B.~Refsdal, M.~Servillat, O.~Streicher, Astron.
  Astrophys. \textbf{558}, A33 (2013)

\bibitem{developers2010networkx}
A.A. Hagberg, D.A. Schult, P.J. Swart, \emph{Exploring network structure,
  dynamics, and function using {NetworkX}}, in \emph{Proceedings of the 7th
  Python in Science Conference}, edited by G.~Varoquaux, T.~Vaught, J.~Millman
  (Pasadena, CA USA, 2008)

\bibitem{pedregosa2011scikit}
F.~{Pedregosa Gael Varoquaux Alexandre Gramfort Vincent Michel Bertrand
  Thirion}, O.~Grisel, {Blondel Mathieu}, P.~Prettenhofer, R.~Weiss,
  V.~Dubourg, J.~Vanderplas, A.~Passos, D.~Cournapeau, M.~Brucher, M.~{Perrot
  Edouard Duchesnay}, J. Mach. Learn. Res. \textbf{12}, 2825 (2011)

\bibitem{paszke2019pytorch}
A.~Paszke, S.~Gross, F.~Massa, A.~Lerer, J.~{Bradbury Google}, G.~Chanan,
  T.~Killeen, Z.~Lin, N.~Gimelshein, L.~Antiga, A.~Desmaison, A.K. Xamla,
  E.~Yang, Z.~Devito, M.~{Raison Nabla}, A.~Tejani, S.~Chilamkurthy, Q.~Ai,
  B.~Steiner, L.F. Facebook, J.B. Facebook, S.~Chintala, \emph{Pytorch: An
  Imperative Style, High-performance Deep Learning Library}, in \emph{Advances
  in Neural Information Processing Systems} (2019), pp. 8026--8037

\bibitem{muller2015python}
E.~Muller, J.A. Bednar, M.~Diesmann, M.O. Gewaltig, M.~Hines, A.P. Davison,
  Front. Neuroinform. \textbf{9}, 11 (2015)

\bibitem{stimberg2019brian}
M.~Stimberg, R.~Brette, D.F. Goodman, Elife \textbf{8}, e47314 (2019)

\bibitem{eppler2009pynest}
J.M. Eppler, M.~Helias, E.~Muller, M.~Diesmann, M.O. Gewaltig, Front.
  Neuroinform. \textbf{2}, 12 (2009)

\bibitem{hines2009neuron}
M.~Hines, A.P. Davison, E.~Muller, Front. Neuroinform. \textbf{3}, 1 (2009)

\bibitem{plotnikov2016nestml}
D.~Plotnikov, B.~Rumpe, I.~Blundell, T.~Ippen, J.M. Eppler, A.~Morrison, arXiv
  preprint arXiv:1606.02882  (2016)

\bibitem{peyser2015nest}
A.~Peyser, W.~Schenck, \emph{The NEST Neuronal Network Simulator: Performance
  Optimization Techniques for High Performance Computing Platforms}, in
  \emph{Posters Presented at the “Society for Neuroscience Annual Meeting"}
  (2015)

\bibitem{TikRub17}
R.A. Tikidji-Hamburyan, V.~Narayana, Z.~Bozkus, T.A. El-Ghazawi, Front.
  Neuroinform. \textbf{11}, 46 (2017)

\bibitem{hines2004modeldb}
M.L. Hines, T.~Morse, M.~Migliore, N.T. Carnevale, G.M. Shepherd, J. Comput.
  Neurosci. \textbf{17}(1), 7 (2004)

\bibitem{dura2019netpyne}
S.~Dura-Bernal, B.A. Suter, P.~Gleeson, M.~Cantarelli, A.~Quintana,
  F.~Rodriguez, D.J. Kedziora, G.L. Chadderdon, C.C. Kerr, S.A. Neymotin, R.A.
  McDougal, M.~Hines, G.M. Shepherd, W.W. Lytton, Elife \textbf{8}, e44494
  (2019)

\bibitem{Dai2020b}
K.~Dai, S.L. Gratiy, Y.N. Billeh, R.~Xu, B.~Cai, N.~Cain, A.E. Rimehaug, A.J.
  Stasik, G.T. Einevoll, S.~Mihalas, C.~Koch, A.~Arkhipov, PLOS Comput. Biol.
  \textbf{16}(11), e1008386 (2020), ISSN 1553-7358

\bibitem{GraBil18}
S.L. Gratiy, Y.N. Billeh, K.~Dai, C.~Mitelut, D.~Feng, N.W. Gouwens, N.~Cain,
  C.~Koch, C.A. Anastassiou, A.~Arkhipov, PLoS ONE \textbf{13}(8), e0201630
  (2018)

\bibitem{Blundell2018}
I.~Blundell, R.~Brette, T.A. Cleland, T.G. Close, D.~Coca, A.P. Davison,
  S.~Diaz-Pier, C.~{Fernandez Musoles}, P.~Gleeson, D.F.M. Goodman, M.~Hines,
  M.W. Hopkins, P.~Kumbhar, D.R. Lester, B.~Marin, A.~Morrison,
  E.~M{\"{u}}ller, T.~Nowotny, A.~Peyser, D.~Plotnikov, P.~Richmond, A.~Rowley,
  B.~Rumpe, M.~Stimberg, A.B. Stokes, A.~Tomkins, G.~Trensch, M.~Woodman, J.M.
  Eppler, Front. Neuroinform. \textbf{12}, 68 (2018)

\bibitem{Gutzen2018}
R.~Gutzen, M.~von Papen, G.~Trensch, P.~Quaglio, S.~Gr{\"{u}}n, M.~Denker,
  Front. Neuroinform. \textbf{12}, 90 (2018)

\bibitem{Mikowski2018}
M.~Mi{\l}kowski, W.M. Hensel, M.~Hohol, J. Comput. Neurosci. \textbf{45}(3),
  163 (2018)

\bibitem{crook2020rep}
S.M. Crook, A.P. Davison, R.A. McDougal, H.E. Plesser, Front. Neuroinf.
  \textbf{14}, 23 (2020)

\bibitem{McdMor17}
R.A. McDougal, T.M. Morse, T.~Carnevale, L.~Marenco, R.~Wang, M.~Migliore, P.L.
  Miller, G.M. Shepherd, M.L. Hines, J. Comput. Neurosci. \textbf{42}(1), 1
  (2017)

\bibitem{RouHin17}
N.P. Rougier, K.~Hinsen, F.~Alexandre, T.~Arildsen, L.A. Barba,
  A.~C.Y.Benureau, C.T. Brown, P.~DeBuy, O.~Caglayan, A.P. Davison, M.A.
  Delsuc, G.~Detorakis, A.K. Diem, D.~Drix, P.~Enel, B.~Girard, O.~Guest, M.G.
  Hall, R.N. Henriques, X.~Hinaut, K.S. Jaron, M.~Khamassi, A.~Klein,
  T.~Manninen, P.~Marchesi, D.~McGlinn, C.~Metzner, O.~Petchey, H.E. Plesser,
  T.~Poisot, K.~Ram, Y.~Ram, E.~Roesch, C.~Rossant, V.~Rostami, A.~Shifman,
  J.~Stachelek, M.~Stimberg, F.~Stollmeier, F.~Vaggi, G.~Viejo, J.~Vitay, A.E.
  Vostinar, R.~Yurchak, T.~Zito, PeerJ Comput. Sci. \textbf{3}, e142 (2017)

\bibitem{Izhikevich2008}
E.M. Izhikevich, G.M. Edelman, Proc. Natl. Acad. Sci. USA \textbf{105}, 3593
  (2008)

\bibitem{izhikevichsite}
E.M. Izhikevich, Personal website, accessed on October, 8th  (2020)

\bibitem{Tuckwell1988}
H.C. Tuckwell, \emph{Introduction to Theoretical Neurobiology: Vol. 1, Linear
  Cable Theory and Dendritic Structure} (Cambridge University Press, Cambridge,
  UK, 1988)

\bibitem{Izhikevich2007}
E.M. Izhikevich, \emph{Dynamical Systems in Neuroscience} (The MIT Press,
  Cambridge, Massachussetts, USA, 2007)

\bibitem{Teeter2018}
C.~Teeter, R.~Iyer, V.~Menon, N.~Gouwens, D.~Feng, J.~Berg, A.~Szafer, N.~Cain,
  H.~Zeng, M.~Hawrylycz, C.~Koch, S.~Mihalas, Nat. Commun. \textbf{9}, 709
  (2018)

\bibitem{Ecker2020}
A.~Ecker, A.~Romani, S.~S{\'{a}}ray, S.~K{\'{a}}li, M.~Migliore, J.~Falck,
  S.~Lange, A.~Mercer, A.M. Thomson, E.~Muller, M.W. Reimann, S.~Ramaswamy,
  Hippocampus \textbf{30}(11), 1129 (2020)

\bibitem{Hendrickson2012}
P.J. Hendrickson, G.J. Yu, B.S. Robinson, D.~Song, T.W. Berger, \emph{Towards a
  large-scale biologically realistic model of the hippocampus}, in
  \emph{Proceedings of the Annual International Conference of the IEEE
  Engineering in Medicine and Biology Society, EMBS} (NIH Public Access, 2012),
  Vol. 2012, pp. 4595--4598

\bibitem{Migliore2014}
M.~Migliore, F.~Cavarretta, M.L. Hines, G.M. Shepherd, Front. Comput. Neurosci.
  \textbf{8}, 50 (2014)

\bibitem{Migliore2015}
M.~Migliore, F.~Cavarretta, A.~Marasco, E.~Tulumello, M.L. Hines, G.M.
  Shepherd, J.G. Hildebrand, P. Natl. Acad. Sci. USA \textbf{112}, 8499 (2015)

\bibitem{Hass2016}
J.~Hass, L.~Hert{\"{a}}g, D.~Durstewitz, PLoS Comput. Biol. \textbf{12},
  e1004930 (2016)

\bibitem{Hagen2016}
E.~Hagen, D.~Dahmen, M.L. Stavrinou, H.~Lind{\'{e}}n, T.~Tetzlaff, S.J. van
  Albada, S.~Gr{\"{u}}n, M.~Diesmann, G.T. Einevoll, Cereb. Cortex \textbf{26},
  4461 (2016)

\bibitem{Senk2018}
J.~Senk, E.~Hagen, S.J. van Albada, M.~Diesmann, arXiv p. 1805.10235 (2018)

\bibitem{Girardi2016b}
G.S. Bortolotto, M.~Girardi{-}Schappo, J.J. Gonsalves, L.T. Pinto, M.H.R.
  Tragtenberg, J. Phys. Conf. Ser. \textbf{686}(1), 012008 (2016)

\bibitem{Arkhipov2018}
A.~Arkhipov, N.W. Gouwens, Y.N. Billeh, S.~Gratiy, R.~Iyer, Z.~Wei, Z.~Xu,
  R.~Abbasi-Asl, J.~Berg, M.~Buice, N.~Cain, N.~da~Costa, S.~de~Vries,
  D.~Denman, S.~Durand, D.~Feng, T.~Jarsky, J.~Lecoq, B.~Lee, L.~Li,
  S.~Mihalas, G.K. Ocker, S.R. Olsen, R.C. Reid, G.~Soler-Llavina, S.A.
  Sorensen, Q.~Wang, J.~Waters, M.~Scanziani, C.~Koch, PLoS Comput. Biol.
  \textbf{14}, e1006535 (2018)

\bibitem{Reimann2013}
M.W. Reimann, C.A. Anastassiou, R.~Perin, S.L. Hill, H.~Markram, C.~Koch,
  Neuron \textbf{79}(2), 375 (2013)

\bibitem{Ramaswamy2018}
S.~Ramaswamy, C.~Colangelo, H.~Markram, Front. Neural Circuit. \textbf{12}, 77
  (2018)

\bibitem{Nolte2019}
M.~Nolte, E.~Gal, H.~Markram, M.W. Reimann, Net. Neurosci. \textbf{4}(1), 292
  (2019)

\bibitem{Tomsett2015}
R.J. Tomsett, M.~Ainsworth, A.~Thiele, M.~Sanayei, X.~Chen, M.A. Gieselmann,
  M.A. Whittington, M.O. Cunningham, M.~Kaiser, Brain Struct. Funct.
  \textbf{220}, 2333 (2015)

\bibitem{Migliore2018}
R.~Migliore, C.A. Lupascu, L.L. Bologna, A.~Romani, J.D. Courcol, S.~Antonel,
  W.A.H. Van~Geit, A.M. Thomson, A.~Mercer, S.~Lange, J.~Falck, C.A. Rössert,
  Y.~Shi, O.~Hagens, M.~Pezzoli, T.F. Freund, S.~Kali, E.B. Muller,
  F.~Schürmann, H.~Markram, M.~Migliore, PLoS Comput. Biol. \textbf{14}(9), 1
  (2018)

\bibitem{TsodyksMarkram1997}
M.~Tsodyks, H.~Markram, P. Natl. Acad. Sci. USA \textbf{94}, 719 (1997)

\bibitem{Hillman1979}
D.~Hillman, \emph{Neuronal Shape Parameters and Substructures as a Basis of
  Neuronal Form}, in \emph{The Neurosciences, Fourth Study Program}, edited by
  S.~F. (Cambridge, MA, USA, 1979), pp. 477--498

\bibitem{Yu2012}
G.J. Yu, B.S. Robinson, P.J. Hendrickson, D.~Song, T.W. Berger,
  \emph{Implementation of topographically constrained connectivity for a
  large-scale biologically realistic model of the hippocampus}, in
  \emph{Proceedings of the Annual International Conference of the IEEE
  Engineering in Medicine and Biology Society, EMBS} (2012), pp. 1358--1361,
  ISBN 9781424441198

\bibitem{Yu2015}
G.J. Yu, P.J. Hendrickson, D.~Song, T.W. Berger, \emph{Topography-dependent
  spatio-temporal correlations in the entorhinal-dentate-{CA}3 circuit in a
  large-scale computational model of the {R}at {H}ippocampus}, in
  \emph{Proceedings of the Annual International Conference of the IEEE
  Engineering in Medicine and Biology Society, EMBS} (Institute of Electrical
  and Electronics Engineers Inc., 2015), Vol. 2015-Novem, pp. 3965--3968, ISBN
  9781424492718

\bibitem{Hendrickson2015}
P.J. Hendrickson, G.J. Yu, D.~Song, T.W. Berger, \emph{A million-plus neuron
  model of the hippocampal dentate gyrus: {D}ependency of spatio-temporal
  network dynamics on topography}, in \emph{Proceedings of the Annual
  International Conference of the IEEE Engineering in Medicine and Biology
  Society, EMBS} (Institute of Electrical and Electronics Engineers Inc.,
  2015), Vol. 2015-Novem, pp. 4713--4716, ISBN 9781424492718

\bibitem{Carnevale2006}
N.T. Carnevale, M.L. Hines, \emph{The {NEURON} Book} (Cambridge University
  Press, Cambridge, Massachussetts, USA, 2006)

\bibitem{yu2013sparse}
Y.~Yu, T.S. McTavish, M.L. Hines, G.M. Shepherd, C.~Valenti, M.~Migliore, PLoS
  Comput. Biol. \textbf{9}(3), e1003014 (2013)

\bibitem{Hertag2012}
L.~Hert{\"{a}}g, J.~Hass, T.~Golovko, D.~Durstewitz, Front. Comput. Neurosci.
  \textbf{6}, 62 (2012)

\bibitem{Douglas1989}
R.J. Douglas, K.~Martin, D.~Whitteridge, Neural Comput. \textbf{1}, 480 (1989)

\bibitem{Girardi2019}
M.~Girardi{-}{S}chappo, M.H.R. Tragtenberg, Phys. Lett. A \textbf{383(36)},
  126031 (2019)

\bibitem{Girardi2020bal}
M.~Girardi{-}{S}chappo, L.~Brochini, A.A. Costa, T.T.A. Carvalho, O.~Kinouchi,
  Phys. Rev. Research \textbf{2}, 012042(R) (2020)

\bibitem{Girardi2021theory}
M.~Girardi-Schappo, E.F. Galera, T.T.A. Carvalho, L.~Brochini, N.L. Kamiji,
  A.C. Roque, O.~Kinouchi, bioRxiv p. 423201 (2020)

\bibitem{Gouwens2018}
N.W. Gouwens, J.~Berg, D.~Feng, S.A. Sorensen, H.~Zeng, M.J. Hawrylycz,
  C.~Koch, A.~Arkhipov, Nat. Commun. \textbf{9}(1), 1 (2018)

\bibitem{Siegle2019}
J.~Siegle, X.~Jia, S.~Durand, S.~Gale, C.~Bennett, N.~Graddis, G.~Heller,
  T.~Ramirez, H.~Choi, J.~Luviano, P.~Groblewski, R.~Ahmed, A.~Arkhipov,
  A.~Bernard, Y.~Billeh, D.~Brown, M.~Buice, N.~Cain, S.~Caldejon, L.~Casal,
  A.~Cho, M.~Chvilicek, T.~Cox, K.~Dai, D.~Denman, S.~de~Vries, R.~Dietzman,
  L.~Esposito, C.~Farrell, D.~Feng, J.~Galbraith, M.~Garrett, E.~Gelfand,
  N.~Hancock, J.~Harris, R.~Howard, B.~Hu, R.~Hytnen, R.~Iyer, E.~Jessett,
  K.~Johnson, I.~Kato, J.~Kiggins, S.~Lambert, J.~Lecoq, P.~Ledochowitsch, J.H.
  Lee, A.~Leon, Y.~Li, E.~Liang, F.~Long, K.~Mace, J.~Melchior, D.~Millman,
  T.~Mollenkopf, C.~Nayan, L.~Ng, K.~Ngo, T.~Nguyen, P.~Nicovich, K.~North,
  G.K. Ocker, D.~Ollerenshaw, M.~Oliver, M.~Pachitariu, J.~Perkins, M.~Reding,
  D.~Reid, M.~Robertson, K.~Ronellenfitch, S.~Seid, C.~Slaughterbeck,
  M.~Stoecklin, D.~Sullivan, B.~Sutton, J.~Swapp, C.~Thompson, K.~Turner,
  W.~Wakeman, J.~Whitesell, D.~Williams, A.~Williford, R.~Young, H.~Zeng,
  S.~Naylor, J.~Phillips, R.C. Reid, S.~Mihalas, S.~Olsen, C.~Koch, bioRxiv p.
  805010 (2019)

\bibitem{Gratiy2018}
S.L. Gratiy, Y.N. Billeh, K.~Dai, C.~Mitelut, D.~Feng, N.W. Gouwens, N.~Cain,
  C.~Koch, C.A. Anastassiou, A.~Arkhipov, PLoS ONE \textbf{13} (2018)

\bibitem{Gewaltig2007}
M.O. Gewaltig, M.~Diesmann, Scholarpedia \textbf{2}(4), 1430 (2007)

\bibitem{Dai2020}
K.~Dai, J.~Hernando, Y.N. Billeh, S.L. Gratiy, J.~Planas, A.P. Davison,
  S.~Dura-Bernal, P.~Gleeson, A.~Devresse, B.K. Dichter, M.~Gevaert, J.G. King,
  W.A.H. {Van Geit}, A.V. Povolotsky, E.~Muller, J.D. Courcol, A.~Arkhipov,
  PLoS Comput. Biol. \textbf{16}(2), e1007696 (2020)

\bibitem{Reimann2015}
M.W. Reimann, J.G. King, E.B. Muller, S.~Ramaswamy, H.~Markram, Front. Comput.
  Neurosci. \textbf{9}, 00120 (2015)

\bibitem{Ramaswamy2015}
S.~Ramaswamy, J.D. Courcol, M.~Abdellah, S.R. Adaszewski, N.~Antille,
  S.~Arsever, G.~Atenekeng, A.~Bilgili, Y.~Brukau, A.~Chalimourda, G.~Chindemi,
  F.~Delalondre, R.~Dumusc, S.~Eilemann, M.E. Gevaert, P.~Gleeson, J.W. Graham,
  J.B. Hernando, L.~Kanari, Y.~Katkov, D.~Keller, J.G. King, R.~Ranjan, M.W.
  Reimann, C.~R{\"{o}}ssert, Y.~Shi, J.C. Shillcock, M.~Telefont, W.~{Van
  Geit}, J.~{Villafranca Diaz}, R.~Walker, Y.~Wang, S.M. Zaninetta,
  J.~DeFelipe, S.L. Hill, J.~Muller, I.~Segev, F.~Sch{\"{u}}rmann, E.B. Muller,
  H.~Markram, Front. Neural Circuit. \textbf{9}, 44 (2015)

\bibitem{Mainen1996}
Z.F. Mainen, T.J. Sejnowski, Nature \textbf{382}, 363 (1996)

\bibitem{Bush1993}
P.C. Bush, T.J. Sejnowski, J. Neurosci. Meth. \textbf{46}, 159 (1993)

\bibitem{BreRud07}
R.~Brette, M.~Rudolph, T.~Carnevale, M.~Hines, D.~Beeman, J.M. Bower,
  M.~Diesmann, A.~Morrison, P.H. Goodman, F.C. Harris, M.~Zirpe,
  T.~Natschl{\"{a}}ger, D.~Pecevski, B.~Ermentrout, M.~Djurfeldt, A.~Lansner,
  O.~Rochel, T.~Vieville, E.~Muller, A.P. Davison, S.~{El Boustani},
  A.~Destexhe, J. Comput. Neurosci. \textbf{23}(3), 349 (2007)

\bibitem{blohm2020}
G.~Blohm, K.P. Kording, P.R. Schrater, eNeuro \textbf{7}(1) (2020)

\bibitem{HofJac17}
F.~H{\"o}fflin, A.~Jack, C.~Riedel, J.~Mack-Bucher, J.~Roos, C.~Corcelli,
  C.~Schultz, P.~Wahle, M.~Engelhardt, Front. Cell. Neurosci \textbf{11}, 332
  (2017)

\end{thebibliography}

\end{document}